%% file: pks_SV.tex
\documentclass[a4paper]{aa}  
\usepackage{graphicx}
\usepackage{epsfig}
\usepackage{color}
\usepackage{txfonts}
\usepackage{natbib,twoopt}
\usepackage{floatrow}
\bibpunct{(}{)}{;}{a}{}{,}

\newcommand{\src}{PKS\,0529-549}
\newcommand{\xs}{X-SHOOTER}

\newcommand{\Htwo}{\mbox{H$_{2}$}}
\newcommand{\Mstar}{\mbox{M$_{\star}$}}
\newcommand{\MHtwo}{\mbox{M$_\mathrm{H_2}$}}
\newcommand{\Msun}{\mbox{M$_{\odot}$}}
\newcommand{\Zsun}{\mbox{$Z_{\odot}$}}
\newcommand{\La}{\mbox{Ly$\alpha$}}
\newcommand{\Ha}{\mbox{H$\alpha$}}
\newcommand{\Hb}{\mbox{H$\beta$}}
\newcommand{\HII}{\ion{H}{II}}

\newcommand{\HeII}{\ion{He}{II}}

\newcommand{\CI}{\mbox{[C\,{\sc i}]}}
\newcommand{\CIonezero}{\mbox{[C\,{\sc i}]\,$^3$P$_{1}$-$^3$P$_{0}$}}
\newcommand{\CItwoone}{\mbox{[C\,{\sc i}]\,(2--1)}}
\newcommand{\CIlevels}{\mbox{[C\,{\sc i}]\,$^3$P$_{2}$-$^3$P$_{1}$}}
\newcommand{\XCI}{\mbox{$X_{\mathrm{[C\,{\textsc{i}}]}}$}}
\newcommand{\LIR}{\mbox{L$_\mathrm{IR}$}}

\newcommand{\Lsun}{\mbox{L$_{\odot}$}}

\newcommand{\SV}{\ion{S}{V}}
\newcommand{\NIII}{\ion{N}{III}}
\newcommand{\NIV}{\ion{N}{IV}}

\newcommand{\NV}{\ion{N}{V}}

\newcommand{\SiII}{\ion{Si}{II}}
\newcommand{\SiIII}{\ion{Si}{III}}
\newcommand{\SiIIIsf}{\ion{Si}{III]}}
\newcommand{\SiIIIf}{\mbox{[Si\,{\sc iii}]}}
\newcommand{\SiIV}{\ion{Si}{IV}}
\newcommand{\CII}{\ion{C}{II}}
\newcommand{\CIIsf}{\ion{C}{II]}}
\newcommand{\CIII}{\ion{C}{III}}
\newcommand{\CIIIsf}{\ion{C}{III]}}
\newcommand{\CIIIf}{\mbox{[C\,{\sc iii}]}}
\newcommand{\CIV}{\ion{C}{IV}}
\newcommand{\OI}{\ion{O}{I}}
\newcommand{\OIIIsf}{\ion{O}{III]}}
\newcommand{\OIV}{\ion{O}{IV}}
\newcommand{\OV}{\ion{O}{V}}
\newcommand{\OII}{\mbox{[O\,{\sc ii}]}}
\newcommand{\OIII}{\mbox{[O\,{\sc iii}]}}

\newcommand{\MgII}{\ion{Mg}{II}}
\newcommand{\FeII}{\ion{Fe}{II}}

\newcommand{\FeV}{\ion{Fe}{V}}

\newcommand{\kms}{\mbox{km\,s$^{-1}$}}
\newcommand{\um}{$\mu\mathrm{m}$}
\newcommand{\dlum}{\mbox{d$_{\mathrm{L}}$}}
\newcommand{\LUV}{\mbox{L$_\textrm{1500\AA}$}}
\newcommand{\Myr}{\mbox{M$_{\sun}$\,yr$^{-1}$}}
\newcommand{\ergs}{erg\,s$^{-1}$}
\newcommand{\betaUV}{$\beta_\textrm{1500\,\AA}$}
\newcommand{\fgas}{\mbox{f$_{\mathrm{gas}}$} }
\newcommand{\tdepl}{\mbox{$t_{\mathrm{depl}}$}}
\newcommand{\eff}{\mbox{$\varepsilon_{\mathrm{ff}}$}}
\newcommand{\fsf}{\mbox{f$_{\mathrm{sf}}$}}
\newcommand{\zsys}{\mbox{$z_{\mathrm{sys}}$}}
\newcommand{\vterm}{\mbox{$v_{\mathrm{terminal}}$}}
\def\la{\mathrel{\hbox{\rlap{\hbox{\lower4pt\hbox{$\sim$}}}\hbox{$<$}}}}
\def\ga{\mathrel{\hbox{\rlap{\hbox{\lower4pt\hbox{$\sim$}}}\hbox{$>$}}}}
\newcommand*{\ditto}{---\textquotedbl---}

\begin{document} 

\title{Quenching by gas compression and consumption}
\subtitle{A case study of a massive radio galaxy at $z=2.57$}
\author{Allison W. S. Man \inst{1,2}\thanks{email:\,allison.man@dunlap.utoronto.ca,\,allisonmanws@gmail.com}, 
Matthew D. Lehnert \inst{3},
Jo\"el D. R. Vernet \inst{1},
Carlos De Breuck \inst{1},
\and
Theresa Falkendal \inst{1,3}
}
          
\institute{European Southern Observatory, Karl-Schwarzschild-Str.\,2, Garching\,bei\,M\"unchen, D-85748, Germany
\and
Dunlap Institute for Astronomy \& Astrophysics, 50 St. George Street, Toronto, ON M5S 3H4, Canada
\and
Sorbonne Universit\'{e}, CNRS UMR 7095, Institut d'Astrophysique de Paris, 98 bis bd Arago, 75014 Paris, France}

\date{Received 30 October 2018; accepted 22 February 2019}

\abstract{
The objective of this work is to study how active galactic nuclei (AGN) influence star formation in host galaxies.
We present a detailed investigation of the star-formation history and conditions of a $z=2.57$ massive radio galaxy based on VLT/\xs\ and ALMA observations.
The deep rest-frame ultraviolet spectrum contains photospheric absorption lines and wind features indicating the presence of OB-type stars.
The most significantly detected photospheric features are used to characterize the recent star formation:
neither instantaneous nor continuous star-formation history is consistent with the relative strength of the \SiII\,$\lambda$1485 and \SV\,$\lambda$1502 absorption.
Rather, at least two bursts of star formation took place in the recent past,
at $6^{+1}_{-2}$\,Myr and $\gtrsim20$\,Myr ago, respectively.
We deduce a molecular \Htwo\ gas mass of $(3.9\pm1.0)\times10^{10}$\,\Msun\ based on ALMA observations of the \CIlevels\ emission.
The molecular gas mass is only 13\% of its stellar mass.
Combined with its high star-formation rate of ($1020^{+190}_{-170}$)\,\Myr,
this implies a high star-formation efficiency of $(26\pm8$)\,Gyr$^{-1}$ and a short depletion time of $(38\pm12)$\,Myr.
We attribute the efficient star formation to compressive gas motions in order to explain the modest velocity dispersions ($\leqslant 55$ \kms) of the photospheric lines and of the star-forming gas traced by \CI.
Because of the likely very young age of the radio source, our findings suggest that vigorous star formation consumes much of the gas and 
works in concert with the AGN to remove any residual molecular gas,
and eventually quenching star formation in massive galaxies. 
}

\titlerunning{Quenching by gas compression and consumption}
\authorrunning{Man, Lehnert, Vernet et al. 2019}
   

\maketitle

\section{Introduction}\label{sec:intro}

Over the past 10 Gyrs, galaxies with stellar masses greater than $10^{11}$\Msun\ are observed to be predominantly quiescent \citep[e.g.,][]{Davidzon2017}.
This implies that the red sequence of the Hubble tuning fork diagram has been in place since early epochs.
The processes underlying their rapid assembly and the eventual cessation of their star formation are not clearly established.
To reproduce the observed number densities of massive quiescent galaxies and match the stellar mass function,
cosmological simulations routinely include the impact of the mechanical and radiative output of active galactic nuclei (AGN) to
halt their episodes of wide spread star formation \citep{dMatteo2005,Springel2005,
Dubois2012, Sijacki2015, Croton2016, Bower2017}.
Given that the most massive supermassive blackholes reside in the most massive galaxies,
and that their accretion is amongst the most energetic events in the Universe,
AGN certainly have sufficient energy to significantly alter the interstellar medium (ISM) of their host galaxies.  
This ``AGN feedback'' can occur either through pressure from the intense radiation fields generated by AGN, winds generated in the accretion disk surrounding the supermassive blackhole, 
or by direct mechanical interaction of powerful radio jets with the surrounding gas. 
In the case of AGN feedback regulating star formation,
one would expect the star-formation rates (SFR) of galaxies to be inversely correlated to their AGN luminosities.
This is however not apparent in observational studies \citep[e.g.,][]{Rosario2013,Stanley2015}.
Bursts of star formation can also create high-speed winds of outflowing gas \citep[e.g.,][]{Heckman1995,Lehnert1996,Lehnert1999,Zirm2005} further complicating a direct understanding of how AGN may regulate star formation. 

Despite the potentially important role that AGN feedback plays in galaxy evolution, 
how such feedback works in detail is not well-understood or well-constrained observationally. 
Understanding the physics of feedback requires knowledge of the characteristics of the AGN such as luminosity and variability timescale,
as well as the density and temperature distributions of the ISM of the host galaxy \citep[e.g.,][]{Wagner2012,Zubovas2013b}.
These detailed properties of AGN and ISM have only been determined for a handful of nearby galaxies with deep, resolved multi-wavelength observations.
While negative feedback may be expected to be prevalent over long timescales, 
AGN feedback can be also positive 
-- enhancing the star-formation rate or efficiency -- 
over shorter timescales \citep{Silk2005,Silk2013}. 
Molecular gas could be entrained by expanding radio bubbles or condensed in-situ from the wake of the buoyant plasma via thermal instabilities,
and blast waves could compress gas leading to enhanced star formation \citep{Fragile2004,Fragile2017,Gaibler2012,Ishibashi2012,Dugan2014}.
Star formation can also be induced by the jet interacting with pre-existing extra-nuclear gas.
This positive form of AGN feedback is supported by resolved observations which find young stars and gas to be aligned with the radio jets,
such as the nearby Centaurus A, 3C\,285, Minkowski's Object near NGC\,541, central cluster galaxies,
as well as 4C\,41.47 at $z=3.7$ \citep{vBreugel1985,Dey1997,Rejkuba2002,Croft2006,Salome2015,Russell2017a,Russell2017b}.

Several inherent challenges impede a direct inference of causality between the AGN activity and the star formation of host galaxies \citep{Volonteri2015b,Volonteri2015a,Harrison2017}.
Emission of stellar populations dominated by O- and B-stars can appear similar to the direct or scattered light of the AGN.
The difficulty of isolating star formation from other sources of emission is innate to most SFR indicators including the UV continuum emission, recombination lines as well as the reprocessed dust emission in the far-infrared.
Uncertainties in measurements of the extinction further complicates the SFR estimate.
Moreover, blackhole accretion rates fluctuate on timescales of $<1$\,Myr,
whereas star formation is essentially obscured during the first few Myr as newly formed stars are still deeply embedded in their natal molecular clouds. 
Typical SFR tracers are sensitive to stellar population ages of a few to 200\,Myr \citep{Kennicutt2012}.
It is therefore difficult to establish, through observed correlations, how exactly the AGN influence the SFR of host galaxies.

Fortunately, many of these challenges can be mitigated by deep UV spectroscopy.
Photospheric absorption lines in the UV are unambiguous signatures of the hot stellar atmospheres of young stars.
When observed in galaxies, they provide constraints on the recent star-formation history \citep{dMello2000}.
Deep UV spectroscopy has been used to study nearby starburst galaxies using, for example, the Goddard High Resolution Spectrograph on board the \textit{Hubble Space Telescope} \citep{Conti1996,Leitherer1996,Heckman1997b}.
Although in principle rest-frame UV spectra are more readily accessible for high redshift objects as they shift to observed optical wavelengths, 
UV photospheric lines have only been detected in a handful of distant galaxies to date, 
as hours of integration time are required to reach the necessary sensitivity even with 8-meter class telescopes.
Common techniques to improve the signal-to-noise of photospheric lines in distant galaxies are gravitational lensing or stacking \citep{Cabanac2008,Quider2009,Quider2010,DZavadsky2010,Bayliss2014,Steidel2016,Rigby2017}.

The physics of AGN feedback can be better understood by studying the UV absorption line spectra of AGN host galaxies.
High-redshift radio galaxies (HzRGs) are ideal for this purpose,
as they are among the most massive and luminous objects at
each redshift \citep{Miley2008}. Moreover, unlike in quasars (type 1
AGN), their AGN emission does not completely outshine the host galaxy because the
obscuring torus conveniently acts as a natural coronagraph. 
A remarkable example is 4C\,41.17, a radio galaxy at $z=3.8$,
whose UV continuum is dominated by starlight rather than AGN light.
Its UV spectrum contains at least one photospheric feature, namely the \SV\ $\lambda$1502 absorption line \citep{Dey1997}, originating from O stars and early B stars \citep{Walborn1995a}.
The detection of the \SV\ photospheric feature, 
together with the alignment of the star-forming regions along the radio jet axis \citep{Steinbring2014}, 
makes 4C\,41.17 a convincing example of AGN-triggered star formation.
Despite the potential of using UV spectroscopy to study feedback physics,
deep spectroscopy of other distant radio galaxies thus far have only resulted in tentative or no detection of photospheric features \citep{Cimatti1998,Smith2010}.

Here we present the most significant detection of multiple rest-frame UV
photospheric absorption lines in \object{\src}, a radio galaxy at $z=2.57$.
\src\ is a massive galaxy with a stellar mass \Mstar~=~$(3\pm2)\times10^{11}$\,\Msun,
\citep{dBreuck2010} and a star-formation rate of SFR~$=1020^{+190}_{-170}$\,\Myr\
\citep{Falkendal2018}. Its \La\ emission extent \citep[$\sim40$\,kpc; ][]{Roettgering1997}  is well beyond its asymmetric double knot radio structure 
\citep[two knots separated by $\sim10$\,kpc; ][]{Broderick2007}.
The radio lobes are embedded within and aligned with the more extended warm emission line gas traced by \OIII\ and \Ha,
as expected for a bi-conical ionized gas outflow \citep{Nesvadba2017,Lelli2018}.

In addition to using the recent star-formation history to understand AGN feedback, it is
equally important to investigate the cold molecular gas that fuels the on-going star formation.
By comparing the molecular gas properties of \src\ with those of other distant massive galaxies,
we can infer how it evolves as massive galaxies quench their star formation.
Moreover, the star-formation efficiency is the ratio of the cold gas mass and the SFR.
Its inverse, the gas depletion time, is an important quantity as it provides a measure of its relative importance to other on-going processes such as AGN outflows.
Here, we probe the cold gas in \src\ by observing the atomic carbon \CIlevels\ emission line, 
which is a tracer of relatively diffuse low-density \Htwo, using the Atacama Large Millimeter/submillimeter Array (ALMA).

In this paper we present a detailed analysis of the star formation properties of \src.
\S\ref{sec:obsandred} provides an overview of the spectroscopic observations and the data reduction.
\S\ref{sec:analysis} describes the analysis of the \xs\ spectrum, 
including the systemic redshift measurement, line identifications, and the characterization of the recent star formation.
\S\ref{sec:CI} presents the derivation of the molecular gas mass and star-formation efficiency based on the ALMA \CI\ spectrum.
In \S\ref{sec:discussion} we discuss \src\ in context of the entire massive galaxy population, 
addressing outstanding questions like:
How do galaxies form stars and eventually quench?
What role do AGN play in this process?
Our findings are summarized in \S\ref{sec:conclusions}.

Throughout this paper, we assume a \citet{Kroupa2001} initial
mass function (IMF)\footnote{When comparing to SFR and \Mstar\ from literature data derived assuming the \citet{Salpeter1955} IMF, we divide the masses by a factor of 1.5 to convert to \citet{Kroupa2001} IMF following \citet{Madau2014}.}.  A cosmology of $H_{0}$ = 70\,\kms\,Mpc$^{-1}$,
$\Omega_\mathrm{M}$ = 0.3 and $\Omega_{\Lambda}$ = 0.7 is adopted.
We adopt the optical velocity definition.

\section{Observations and Reduction}\label{sec:obsandred}

\subsection{VLT/\xs\ observations} \label{sec:xs_obs}

\begin{figure}[!ht]
\begin{centering}
\includegraphics[width=9cm]{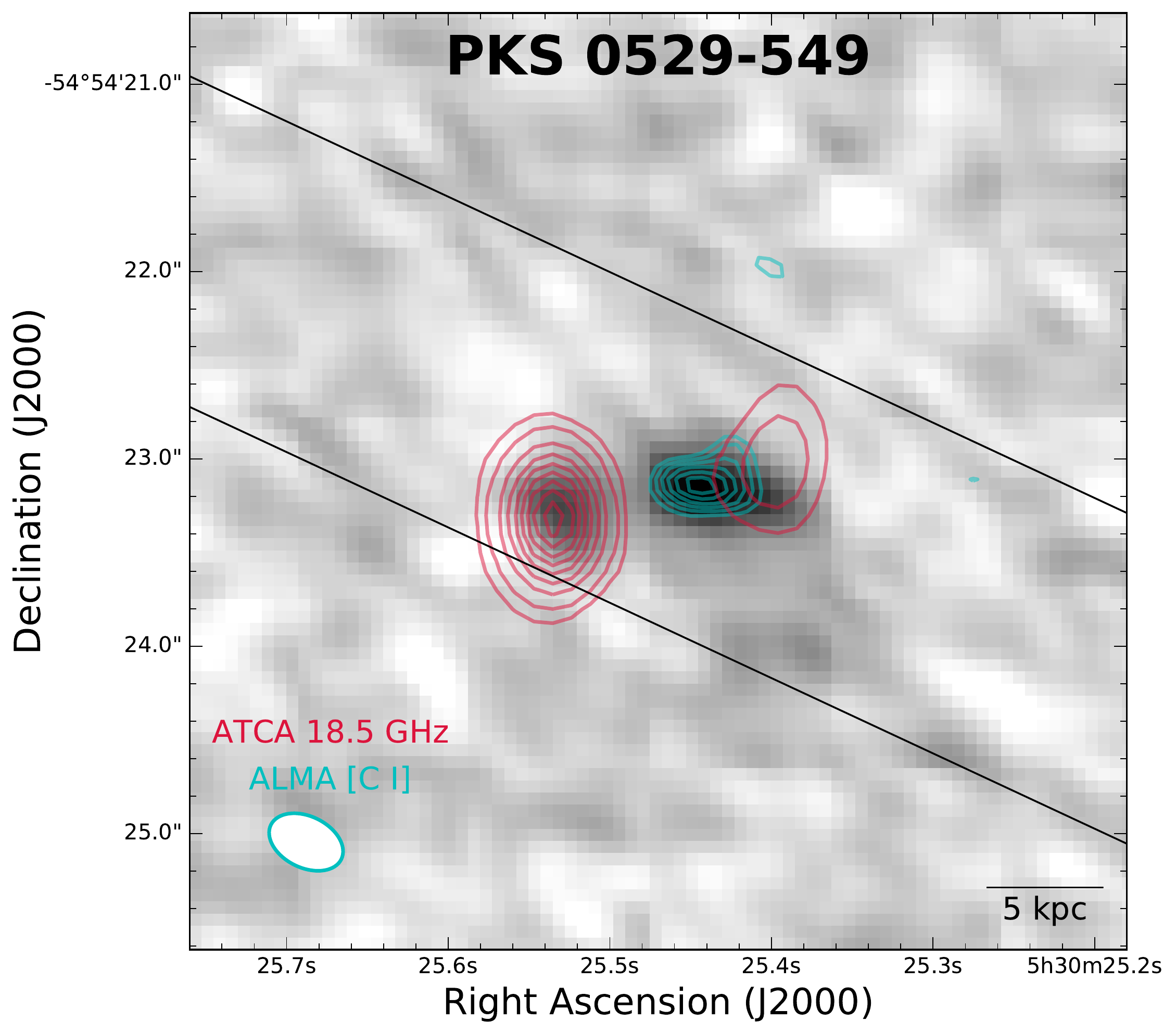}
\caption{
The \xs\ slit is shown as black lines superposed on the greyscale rendition of the ALMA Band 6 continuum image.  
The displayed image is $5\arcsec\times5\arcsec$.  
\CIlevels\ emission is only detected in the host galaxy (western component in the greyscale ALMA continuum image),
as shown in moment-zero map plotted as cyan contours (at 0.3, 0.35, 0.4, 0.45, 0.5, and 0.55 Jy\,beam$^{-1}$\,\kms).
The 18.5\,GHz ATCA radio continuum shows a double lobe structure and is overlaid as red contours (at 1, 2, 4, 6, 8, 10, 12, 14, 16 mJy\,beam$^{-1}$).
\CIlevels\ emission is not detected in any of the radio lobes. 
Note that the eastern radio lobe is detected in the ALMA Band 6 continuum image and is interpreted as pure synchrotron emission \citep{Falkendal2018}.
The beam size of the ALMA image is shown in the lower left corner.
\label{fig:slit}
}
\end{centering}
\end{figure}

Observations were obtained using the \xs\ echelle spectrograph
\citep{Vernet2011} mounted on UT\,3 of the ESO Very Large Telescope through programme 092.B-0772 (PI: M.~Lehnert).  
They were conducted in service mode over five separate nights between 2013 October 30
and 2014 January 29.  
The seeing full-width half-maximum (FWHM) was reported to range from 0$\arcsec$.7--1$\arcsec$.7, 
on average 1$\arcsec$.1.
The science observations were taken in airmass range of 1.16--1.41,
with median (mean) airmass of 1.21 (1.23).
The atmospheric dispersion corrector (ADC) was dysfunctional and therefore disabled during this period.
The \xs\ slit centroid, quoted as the median position of all exposures,
is provided in Table~\ref{table:prop}.
The slit was oriented at a position angle of $-115^{\circ}$ from N to E as shown in Figure~\ref{fig:slit},
near the parallatic angle to minimize slit losses.
Each arm, UVB, VIS, and NIR, had different slit widths, 1$\arcsec$.6,
1$\arcsec$.5, and 1$\arcsec$.2, corresponding to moderate spectral resolutions of
$\lambda/\Delta\lambda\sim$3\,200, 5\,000, and 4\,300,
which were sampled at 8.9, 10.3, and 22.6 pixels per FWHM, respectively.
The slit was broad enough to cover both the host galaxy and the radio lobe of \src\ (corresponding to the W and E components as presented in \citealt{Broderick2007,Falkendal2018}).
The integration times per exposure in the UVB, VIS, and NIR arms were 1\,800, 900, 600\,seconds, respectively.
The dithering strategy was mostly done in NODDING mode, 
with a few exposures done in STARE mode.  
The total on-source exposure time was 8.5\,hours.

\begin{table}
\centering
\caption{Properties of PKS\,0549-529. The position refers to the \xs\ slit centroid. The UV and optical continuum slopes are measured from the observed spectrum.
The quoted \LIR\ refers to that attributed to star formation only, excluding any AGN contribution \citep{Falkendal2018}.}
\label{table:prop}
\begin{tabular}{ll}
\hline
\hline
Property & Value \\
\hline
RA & $5^{h}30^{m}25^{s}.565$ \\
DEC & $-54^{\circ}54'22\arcsec.624$ \\ 
Systemic redshift, \zsys & $2.5725\pm0.0003$ \\
Stellar mass, log(\Mstar/\Msun) & $(3\pm2)\times10^{11}$ \\
UV continuum slope, \betaUV & $-0.27\pm0.05$ \\
Optical continuum slope, $\beta_\textrm{2500\,\AA}$ & $-0.61\pm0.02$ \\
UV luminosity, log(\LUV/\Lsun) & 10.26 $^{+0.14}_{-0.21}$ \\
Infrared luminosity, log(\LIR/\Lsun) & $12.9\pm0.1$ \\
\hline
\end{tabular}
\end{table}

\subsection{\xs\ reduction}\label{subsec:XSred}

We used the EsoRex pipeline version 2.8.19 \citep{Modigliani2010} through the Reflex interface \citep{Freudling2013} to reduce the spectroscopic
data. For optimal sky subtraction, the UVB and VIS arm exposures were
reduced as STARE mode,  while the NIR arm exposures were reduced in
NOD mode. Default pipeline parameters were used, with the following modifications: (1) Median sky subtraction; (2) Rows of sky pixels were
manually defined for each exposure; and (3) a constant sky background
on the raw frames was assumed and fit with a line down each column.

The uncertainty in the wavelength calibration is dominated by the systematic uncertainties of 0.3, 0.2, and 0.04\,\AA~in the UVB, VIS, and NIR arms, respectively,
corresponding to 20\,\kms\ at observed 450\,nm and 0.6\,\kms\ at 2\,\um.
The UVB arm native pixel scale 0.1\,\AA, 
which samples the resolution at our selected slit width over 8.9 pixels, 
corresponds to FWHM of 0.89\,\AA~in the observed frame. 
This implies a velocity resolution of $\sigma_{\mathrm{instr}}$~=~0.1\,\AA~or 20\,\kms\ at a rest-frame wavelength of 1500\,\AA\ for the redshift of \src, $z=2.57$.
The reduced \xs\ data are in topocentric velocity reference frame.
Correction to barycentric velocity frame ranges from $-5.6$ to 3.4\,\kms\ with a standard deviation of 2.5\,\kms\ across exposures. This level of correction is below the spectral resolution.
Correction to barycentric velocity frame was therefore not applied to avoid introducing errors by resampling the spectrum.

To spatially align the exposures across different dither positions, 
we used the brightest emission lines in each arm to determine their centroids.  
For the three arms, we used the blue peak of the \La\
emission at $\sim4336.3$\,\AA, \CIIIsf\ emission at $\sim6815.2$\,\AA,
\OIII\,$\lambda$5008 emission at $\sim1.7889$\,$\mu$m for the UVB, VIS, and NIR arms respectively.
In each exposure, we determine the shift along the spatial direction
by searching for the brightest row summed over 10 spectral pixels.
The combined 2D spectrum is assembled using the median pixel value of different exposures, 
after applying the spatial shift.
The orders were merged manually as the pipeline merged spectra had dips in the fluxes where orders overlap.

To extract the one-dimensional (1D) spectrum, we summed the
flux densities contained in the central nine rows to optimize
the signal-to-noise ratio (S/N) of the absorption line features.
To correct for Galactic reddening, we use the Galactic extinction
map\footnote{\url{http://irsa.ipac.caltech.edu/applications/DUST}}
of \citet{Schlafly2011} to determine a value of E(B--V) = 0.0550 at
the source position.  We correct the spectrum with a reddening curve
of $A_{V}$ = 0.1706 and the extinction curve of \citet{Cardelli1989},
assuming $A_{V}/E(B-V)=3.1$.  

\subsection{ALMA Band 6 observations and reduction}

\src\ was observed with ALMA through the Cycle 2 programme 2013.1.00521.S (PI: C. De Breuck).
The Band 6 observations were conducted on 2014 September 2 with
34 antennas,
with a total on-source integration time of 5\,minutes.
We used four 1.875\,GHz spectral windows, one of which was tuned to
observe the \CIlevels\ (hereafter \CI) line in \src. 
The data were calibrated in the Common Astronomy Software Application \citep[CASA; ][]{McMullin2007} with the supplied calibration script.  To maximize S/N we used natural weighting
and a coarse channel width of 50\,\kms\ in the barycentric velocity frame.
The root-mean-square (rms) noise was 48\,$\mu$Jy in the continuum image,
and 0.5\,mJy at channel width of 50\,\kms\ around the region of the \CI\ emission line.  
The restoring beam used was $0\arcsec.43 \times 0\arcsec.28$ at position angle of $65^{\circ}$.
Full details about the calibration of the ALMA observations are provided in \citet{Falkendal2018} and \citet{Lelli2018}.

\begin{figure*}[!htbp]
\begin{centering}
\includegraphics[width=\linewidth]{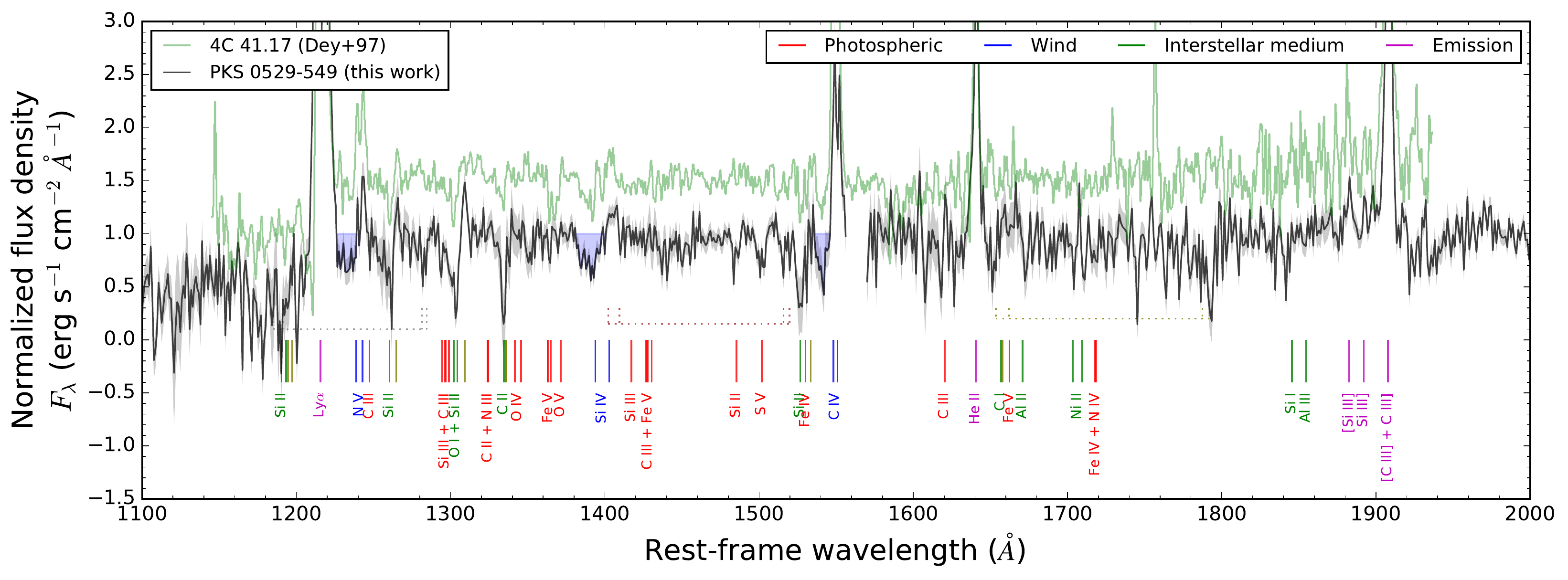}
\caption{
Rest-frame continuum-normalized UV spectra of \src\ (black; this work) and 4C\,41.17 \citep[green;][]{Dey1997}. 
We indicate the wavelengths of the most prominent lines
in the spectra annotated with the ion responsible.  
The colours represent the likely dominant origin of the lines: 
red for stellar photospheric, 
blue for resonant doublets from stellar winds and AGN,
green for low-ionization lines in the interstellar medium,
olive for their associated excited fine-structure emission,
and purple for bright emission lines.
The blue shades show the absorption wings of the resonant lines originating from stellar winds.
The dotted lines indicate possible ISM absorption of \FeII\,$\lambda\lambda$2587,2600 and \MgII\,$\lambda\lambda$2796,2804 at $z=0.64$ (grey), $z=0.94$ (brown), and $z=1.28$ (olive).
For easier visualization, the spectrum of \src\ has been binned by twice the FWHM (i.e., variance-weighted mean for every 18 / 21\,pixels for the UVB / VIS arms, respectively).
The spectrum of 4C\,41.17 has been smoothed by a Gaussian kernel, and shifted upwards by 0.5. 
The grey shade shows the noise as propagated from the \xs\ pipeline without accounting for correlated noise.
The spectrum of \src\ has been shifted to vacuum wavelengths.
The gap at $\sim1550$--$1570$\,\AA\ is the noisy region between the UVB and VIS arms of \xs.
}
\label{fig:spec_hzrg}
\end{centering}
\end{figure*}

\section{Results I: Bursty recent star-formation history}\label{sec:analysis}

\subsection{Systemic redshift}\label{sec:sysred}

We measured redshifts by fitting 
Gaussian profiles to the detected emission lines masking
regions that are strongly influenced by sky line emission.  In case of doublets,
such as \CIV, \CIIIsf, and \OII,
we fit each component of multiplets independently (Table~\ref{table:redshifts}).
The rest-frame wavelengths were taken from the atomic line list v 2.04 maintained by
Peter van Hoof\footnote{\url{http://www.pa.uky.edu/~peter/atomic}\label{footnote:atomicline}} whenever available,
and the rest were drawn from literature \citep{dMello2000,Pettini2000,Leitherer2011,Steidel2016}.
Reassuringly, the redshifts of most emission lines are consistent within
their measurement uncertainties,
with the exception of the resonant wind lines (i.e., \NV, \SiIV, and \CIV; Fig.~\ref{fig:em_velo}).
We therefore define the systemic redshift with the broad \HeII\ line emission (\zsys\ = $2.5725\pm0.0003$), 
and therewith infer the velocity offsets.
The high S/N and spectral resolution of the \xs\ spectrum,
as well as its broad wavelength coverage,
provides a wide range of lines for us to explore the reasons as to why some lines show velocities that are highly discrepant. Here, we simply note the lines with anomalous profiles and/or velocities to inform the discussion later in this paper.

\begin{itemize}

\item The \HeII\ profile is asymmetric and appears to be multiply
peaked (Fig.~\ref{fig:vel_comp}). 
While we expect the AGN scattered broad-line emission to dominate the \HeII\ profile, 
which justifies its use to define the systemic velocity,
in \S\ref{sec:photosphere} and \S\ref{sec:vel_comp} we discuss evidence
that the intense star formation in \src\ might contribute to the complexity of \HeII\ profile.

\item The spatially extended broad \La\ emission component has (surprisingly) consistent velocity with \zsys.  An additional narrower \La\ absorber is
redshifted by 101\,\kms, having line widths which vary across the slit.
This results in an asymmetric double-peaked profile.  

\item The resonant doublets \NV, \SiIV, and \CIV\ are redshifted with
respect to $z_{\mathrm{sys}}$.  This is due to the emission peaks being
affected by blueshifted absorption, as these lines appear to have
P-Cygni profiles.  In \S~\ref{sec:winds}, we discuss the nature of these
resonant lines in more detail.

\end{itemize}

\subsection{Strong stellar photospheric lines}\label{sec:photosphere}

Photospheric absorption lines are often used to characterize the photospheres of the stars that dominate the continuum emission.
When observed in galaxies,
they provide constraints on the star-formation history and metallicity of stellar populations.
They arise from excited, mostly non-resonant, energy levels
and therefore are generally not associated with interstellar gas.
Rest-frame UV continuum emission is typically dominated by hot, massive stars in young populations ($\la$ 100\,Myrs).
In fact, only stars more massive than $5\,\Msun$ (corresponding to spectral type B5V) contribute significantly to the UV spectrum blue-ward of 2200\,\AA\ \citep{Heckman1997b}.

Photospheric features detected in \src\ are listed in Table~\ref{table:photospheric}.
As apparent in Fig.~\ref{fig:spec_hzrg},
the most prominent photospheric absorption lines detected are \SiII\,$\lambda$1485 and \SV\,$\lambda$1502 (excitation potentials of 6.83 and 15.76\,eV, respectively).
\SV\,$\lambda$1502 is observed in most types of O-stars of luminosity classes I--V,
as well as early B-type stars (B0--B1). 
\SiII\,$\lambda$1485, unlike the other ground-state, interstellar Si absorption lines commonly seen in high-redshift galaxies (e.g., \SiII\,$\lambda\lambda\lambda\lambda$1206, 1260, 1304,
1526; \citealt[][and references therein]{Steidel2016}), arise from an excited level in their ions.
Therefore, these lines originates in the photospheres of hot stars -- observed in mid- to late-B stars (B3--B8 and weaker in B-supergiants; \citealt{Rountree1993,dMello2000}) --
only appearing 20\,Myr after a burst of star formation.
The simultaneous appearance of these photospheric absorption lines is key to constrain the recent star-formation history of \src\ in §\ref{sec:sb99}.

While in principle the photospheric
features could provide the best definition of \zsys,
it is not straightforward to accurately determine their redshifts for
the following reason.  Because \xs\ has a moderately high spectral
resolution, both \SiII\,$\lambda$1485 and \SV\,$\lambda$1502 are resolved
into multiplets (Table~\ref{table:photospheric}, Fig.~\ref{fig:vel_comp}). 
None of the observed spectra of nearby starburst galaxies and
individual stars of which we are aware have sufficient spectral resolution to resolve these multiplets for accurate profile fitting. 
Knowledge of the surface gravity (log $g$) and temperature of the stars they arise in are necessary to determine the relative absorption line strengths of these multiplets. 
We note that assuming the \zsys\ as determined from the broad \HeII\ emission line provides a satisfactory match between the stellar features we have observed and the model stellar spectra (\S\ref{sec:sfh}, Fig.~\ref{fig:compare_sb99_multipanel}).

The complex \HeII\,$\lambda$1640 profile in the spectrum of \src\ appears as if it could have redshifted absorption components (Fig.~\ref{fig:vel_comp}; Table~\ref{table:photospheric}). 
Although the broad \HeII\,$\lambda$1640 emission is expected to be dominated by the scattered
emission from the AGN, it is an absorption feature in O- and early B-type
stars, and can be observed in emission and/or absorption (P-Cygni profile)
in Wolf-Rayet stars and O-type supergiants  \citep[Fig.~\ref{fig:spec_OBstars}; ][]{Willis1982,Kinney1993,dMello2000}.
We will further discuss the nature of the complex \HeII\ profile in \S\ref{sec:vel_comp}.

At any rate, we conclude that we have accurately identified stellar photospheric lines as these
absorption features have observed wavelengths consistent with the \zsys\ of \src.

\input{table_photospheric.tex}

\subsection{Identifying weaker photospheric absorption lines}\label{sec:photosphere_weak}

There are other lines in the spectrum that could originate from stellar photospheres, but are blended with nearby features or could have non-stellar origins.
They are listed as follows:

\begin{itemize}

\item \CIII\,$\lambda$1247 is
observed in stars of spectral types O3--O9 and B0--B9, and lies close to the \NV\,$\lambda\lambda$1239, 1243 resonant doublet.
It has a P-Cygni profile which is only seen in hyper-giant stars of types O9.5 Ia and B0 Ia (Fig.~\ref{fig:spec_OBstars}). 

\item \SiII\,$\lambda\lambda\lambda$1260, 1304, 1526 absorption are detected. They are strong features in B-type stars and their presence are expected given the significant detection of \SiII\,$\lambda$1485 (§\ref{sec:photosphere}). 
However, these \SiII\ lines are not particularly constraining as they are ground-state transitions that can also arise from the interstellar medium.
Significant interstellar medium absorption is suggested by the depth of these absorption lines and their blueshifted absorption wings, both of which are not reproduced in model spectra (Fig.~\ref{fig:compare_sb99_instant}).

\item A blend of photospheric lines at
1294--1299\AA~(\SiIII\,$\lambda\lambda\lambda$1294, 1296, 1298,
\CIII\,$\lambda$1296) originating from B-type stars are detected.
They are however affected by the broad ISM \OI\ and \SiII\ absorption line blend at $\sim$1300\,\AA.  

\item A blend of \CII\,$\lambda$1323 and \NIII\,$\lambda$1324 are detected as well. These features are strongest in late B-type stars \citep[see Fig.~8 of][]{dMello2000}.

\item Two line indices 1370 (covering 1360--1380\AA) and 1425 (covering
1415--1435\AA) have been calibrated as metallicity indicators,
as they are not dependent on the age of the young stellar population
\citep{Leitherer2001, Rix2004, Maraston2009}. 
The features \OIV\,$\lambda$1342, \OV\,$\lambda$1371 and the nearby
four \FeV\,$\lambda\lambda\lambda\lambda$1345, 1362, 1363 and 1364 lines fall into
the 1370 index.
The features \SiIII\,$\lambda$1417, \CIII\,$\lambda\lambda$1426, 1427 and
\FeV\,$\lambda\lambda$1427, 1430 fall into the 1425 index.
These photospheric lines are typical of O and early B stars \citep{Leitherer2001}.
The \FeV\ absorption features are strong,
indicating substantial metallicity.

\end{itemize}

Interestingly, a number of fine-structure excited transitions features
are in emission.  
They include \CIII\,$\lambda\lambda$1426, 1428 doublet,
\CIII\,$\lambda$1620, and \SiII\,$\lambda\lambda\lambda$1265, 1309, 1533.
The \CIII\ lines are expected to be in absorption if they originate from the photospheres of O-stars, and in emission in WN and WC stars
\citep{Kinney1993}.  
The \SiII\ lines are expected to be in absorption
in B-stars (\S\ref{sec:sb99}),
and \SiII\,$\lambda$1533 can be seen in P-Cygni in some Wolf-Rayet stars \citep{Willis1982}.
The comparison with stars of various spectral types is shown in Fig.~\ref{fig:spec_OBstars}.
We will discuss their nature further in \S\ref{sec:sb99}.

\subsection{Stellar wind features}\label{sec:winds}

We detect the resonant doublets \NV\,$\lambda$1240, \SiIV\,$\lambda$1400, and \CIV\,$\lambda$1550, which appear to have P-Cygni profiles.
These high-ionization lines are observed in stellar winds of
O- and B-type stars \citep[Figure~\ref{fig:spec_OBstars};][]{dMello2000,
Robert2003}.
In hot stars, these lines are observed to have broad absorption line components with a range of terminal velocities, of hundreds to thousands of \,\kms, varying with spectral type.
\CIV\ is the strongest rest-frame UV stellar feature in O-type stars.  Both \CIV\ and \SiIV\ are absent in stars later than
B5 \citep{dMello2000}.  
\SiIV\ is typically weaker than the two other resonant stellar wind lines \citep{Leitherer2011}.
These stellar wind features are expected given that OB-stars dominate the UV continuum emission,
as suggested by the observed photospheric features discussed in \S\ref{sec:photosphere} and \S\ref{sec:photosphere_weak}. 

The velocity profiles of \NV, \SiIV, and \CIV\ are in accord with being winds.
While the line emission is expected to originate from the broad line region of the AGN \citep{Vernet2001},
the observed blueshifted absorption profiles are indicative of a stellar origin rather than the interstellar medium for several reasons.
First, their ionization energy is high: Ionization potentials of 33.5 and 77.5\,eV are required to create Si$^{+3}$ and N$^{+4}$ ions,
corresponding to the \SiIV\ and \NV\ absorption lines, respectively
\citep{Morton2003}.  An interstellar origin is ruled out, as it is unlikely to have such hard photons and/or densities for collisional excitation. 
Second, the detection of photospheric features of massive stars, e.g., \SiII\,$\lambda$1485 and \SV\,$\lambda$1502 
(\S\ref{sec:photosphere}), means that it is likely that the UV continuum emission is predominately due to massive stars.
Thus we would expect to also observe the absorption of the UV continuum arising in the stellar winds of those same early-type stars that dominate the continuum.
Lastly, their velocity structure is vastly different from the saturated, broad absorption lines without the accompanying redshifted emission seen in gamma-ray burst afterglows \citep{Fox2008} that arise from the circumburst medium.  All these arguments support our interpretation that the observed the broad absorption lines of the resonance doublets of \NV, \SiIV, and \CIV, are due to stellar winds.

What evidence do we have to conclude that the winds arise from hot, massive stars rather than the quasar accretion disk?
Qualitatively, stellar winds and quasar-driven winds are the
same phenomenon: radiation pressure generated by the UV photons
which are resonantly trapped by high column density ionized gas. The photon
trapping extracts the momentum from the UV radiation field which
drives stellar winds and accretion disk winds of highly ionized gas.
There are, however,
differences in physical scales and the structure of
the surrounding medium leading to different line profiles.
Winds from rotationally-supported quasar accretion disk viewed through the outflow are classified as broad absorption line (BAL) quasars \citep{Elvis2000}.  
Models of quasar accretion disk wind predict that high-ionization absorption lines in
BAL quasars are often saturated and at times black \citep{Arav1999a,
Arav1999b, Higginbottom2013, Matthews2016}. 
Observationally, absorption features in BAL quasars are seen to have discrete velocity structure that is indicative of regions
where the outflows collide with the ambient interstellar medium or
narrow-line region gas.
In Fig.~\ref{fig:spec_bal}, it is apparent that the high-ionization absorption lines of \src, 
\NV, \SiIV, and \CIV, 
have notably distinct velocity structures compared to TXS\,J1908+7220, 
a radio galaxy at $z$=3.54 with BAL features \citep{Dey1999, dBreuck2001}. 
In \src\ these wind features appear smooth and do not have discrete velocity components as is observed in accretion disk winds.
Rather, they appear akin to a collection of OB-stars \citep[Fig.~\ref{fig:spec_OBstars};][]{Walborn1985,Walborn1995b}. 
Spectra of BAL quasars lack photospheric features that are abundantly seen in \src\ (\S\ref{sec:photosphere} and \S\ref{sec:photosphere_weak}).
Thus the wind features seen in \src\ do not have the same origin as those in BAL quasars, i.e., they are not quasar accretion disk winds.

Making the reasonable assumption that stellar winds cause the broad absorption lines of
the resonant doublets in \src, 
we can deduce the spectral types of stars contributing to these features.  
We define the terminal velocity, \vterm, as the maximum velocity reached of the blueshifted absorption component,
as illustrated by the blue shades in Figure~\ref{fig:spec_hzrg}.
To estimate the measurement errors in \vterm,
we produce 100 Monte-Carlo realizations of the \xs\ spectrum (0.5\,\AA\ binning) using the error spectrum,
repeat the measurement,
and report the mean and standard deviation of \vterm\ as listed in Table~\ref{table:vterm}.
The \NV, \SiIV, and \CIV\ resonant doublets have absorption components with \vterm~$\sim2600-3000$\,\kms\ with errors of 400--500\,\kms.
Such high terminal velocities are similar to those seen in O-stars over a full range of surface gravities \citep{Prinja1990}.  
We thus infer that O-stars are the main contributor to the winds observed in \src.
B supergiants are unlikely to contribute to the stellar winds observed here, given that their winds are characterized by lower terminal velocities,
\vterm~$<$~200\,\kms. The characteristics of the broad absorption lines are consistent
with very young, a few Myr old, massive stars in \src\ as we will elaborate in \S\ref{sec:sb99}.

\input{table_vterm.tex}

\subsection{Characterizing the young stellar population}\label{sec:sfh}

Having identified signatures of young stars in the UV spectrum of \src\ in the previous subsections,
we now proceed to quantify its recent star-formation history by comparing to model spectra and observed spectra of galaxies.

\subsubsection{Evidence for recent bursts of star formation}\label{sec:sb99}

\begin{figure}[!ht]
\begin{centering}
\includegraphics[width=\linewidth]{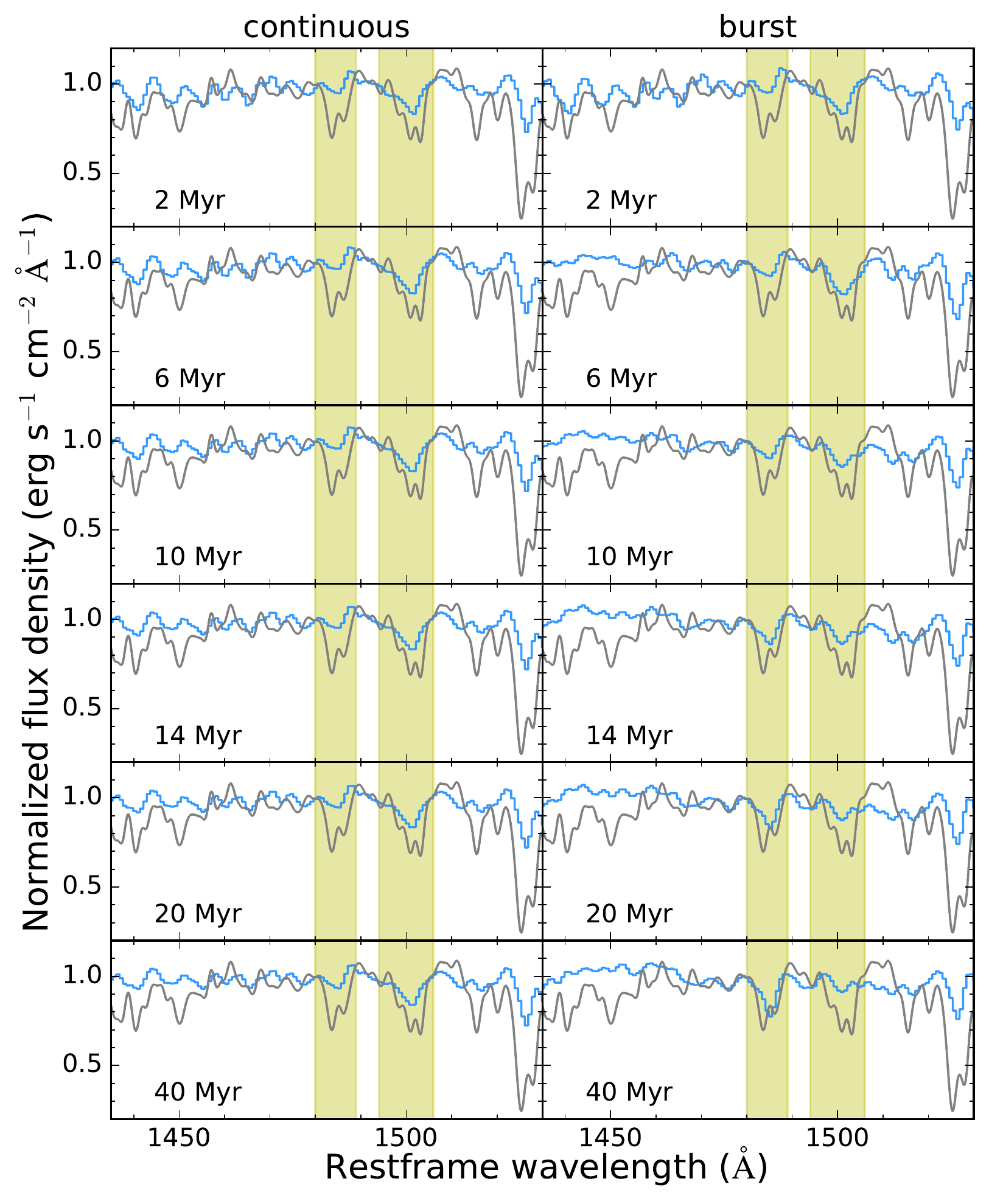}
\caption{Comparison of \texttt{Starburst99} model spectra \citep[in blue; ][]{Leitherer1999} to the observed spectrum of \src\ (in grey) over the spectral region, 1435--1540\,\AA. 
Model spectra of continuous star formation are compared with the spectrum of \src\ on the left and models with a single burst of star formation are compared on the right.
We highlight in yellow the spectral regions of containing the two most significantly detected photospheric lines, \SiII\,$\lambda$1485 and \SV\,$\lambda$1502. The ages of individual spectral models are indicated in the lower left of each panel. 
The spectra have been normalized to the continuum. For visualization both the model and observed spectra have been smoothed to matching resolution.}
\label{fig:compare_sb99_multipanel}
\end{centering}
\end{figure}

\begin{figure*}[htbp!]
\begin{centering}
\includegraphics[width=\linewidth]{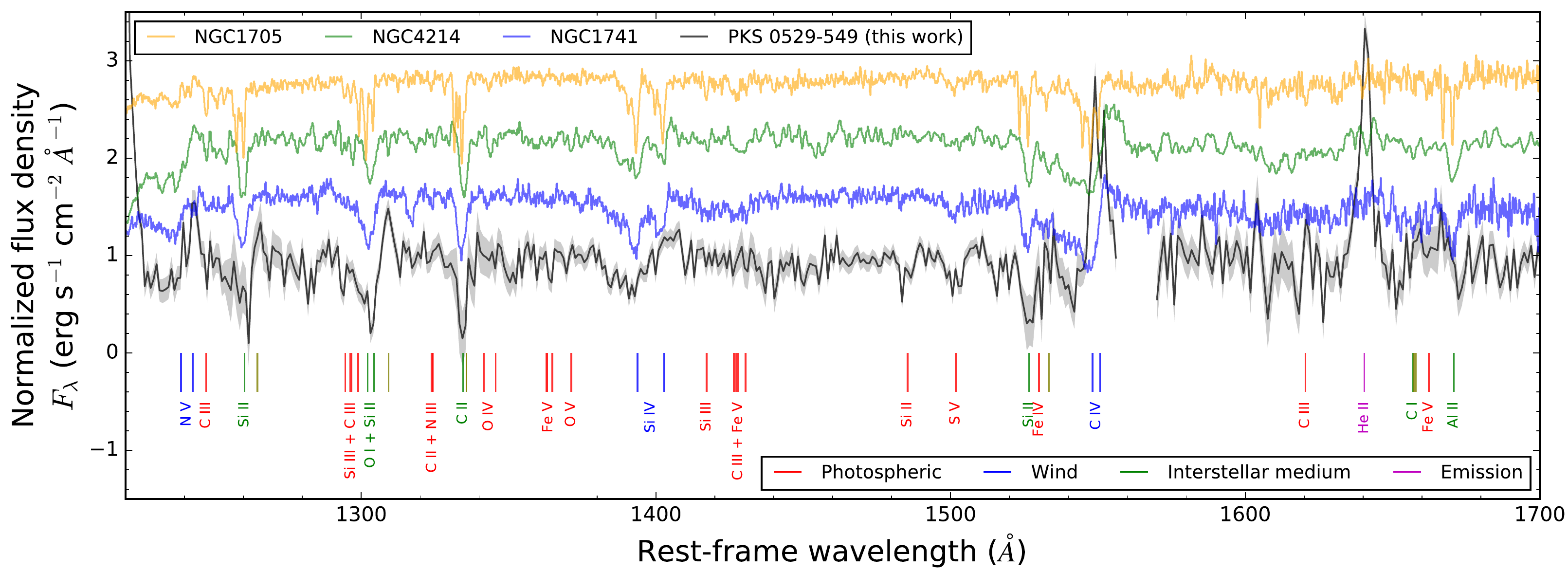}
\caption{
Comparison of the UV spectra of \src\ (black; this work) to three nearby dwarf galaxies with recent star formation.
NGC\,4214-1 (green) and NGC\,1741-B1 (blue) are both aged $\sim4-5$\,Myr,
while NGC\,1705-1 (yellow) is slightly older with age $\sim$10\,Myr \citep{Conti1996,Leitherer1996,Heckman1997b,dMello2000}. 
Following the colour scheme of Figure~\ref{fig:spec_hzrg},
we indicate the wavelengths of various lines in the spectra annotated with the ion responsible,
and show the error spectrum as a grey shade.
Note that the blue absorption edge of \SiIV, seen in all galaxies plotted here except the most evolved NGC\,1705, suggests the presence of a population of O-type supergiants that appear at 3--6\,Myr after an instantaneous burst \citep{Conti1996}.
The spectra have been continuum normalized to unity,
and shifted upwards in increments of 0.6 for visualization.
}
\label{fig:spec_ngc}
\end{centering}
\end{figure*}

To derive constraints on the star-formation history of \src,
we compare the stellar photospheric and wind features we observe to the model
spectra generated with the \texttt{Starburst99} tool v7.0.1 \citep{Leitherer1999,
Leitherer2014}.  As input parameters, we adopt a \citet{Kroupa2001}
initial mass function and the default values for supernova and blackhole
cut-off masses. We use the Geneva stellar evolution tracks with zero
stellar rotation, along with the mass loss rates, at solar metallicity
\Zsun~=~0.014 \citep{Ekstrom2012}.  All other parameters used are the
recommended values.

The star-formation histories used to generate the model spectra probe two extreme cases:
(1) an instantaneous burst creating a stellar population with an initial mass of 10$^{6}$\,\Msun; and
(2) a continuous star-formation rate of 1~\Myr.  
A comparison of these model spectra at different ages to that of \src\ is shown in Figure~\ref{fig:compare_sb99_multipanel},
zoomed to the wavelength range containing the two most significantly detected photospheric features, 
namely \SiII\,$\lambda$1485 and \SV\,$\lambda$1502 having comparable equivalent widths (\S\ref{sec:photosphere}).
In general, the instantaneous burst model provides a better, although still imperfect, match to the photospheric line depths than the continuous star formation model which is unable to reproduce the depth of the \SiII\,$\lambda$1485 absorption.
Further inspection of the \textit{relative} strengths of the two photospheric lines provides clues to a bursty star-formation history.
A younger age of 3--6\,Myr provides the deepest (although still insufficient) \SV\,$\lambda$1502 line,
while an older age of $\gtrsim20$\,Myr is needed to reproduce the depth of the \SiII\,$\lambda$1485 line.
This can be easily understood once the origins of these photospheric features are considered (\S\ref{sec:photosphere}):
\SV\,$\lambda$1502 originates from O- and early B-type stars, 
whereas \SiII\,$\lambda$1485 originates from mid- and late B-type stars.
The simultaneous occurrence of the two features with such depths implies that more than one burst took place in the recent past.
The simplest explanation would be two bursts of star formation took place in the past $\sim$100\,Myr.
To identify the best-fitting model,
we heavily bin and resample the \xs\ spectrum to match the resolution of the model spectra,
and use the associated variance to compute chi-square values in the wavelength ranges of 1480--1489\,\AA\ and 1494--1506\,\AA\,
corresponding to the \SiII\,$\lambda$1485 and \SV\,$\lambda$1502 features.
We determine that the models of a burst age of $6_{-2}^{+1}$\,Myr and $\gtrsim20$\,Myr provide the best fit to the \SV\,$\lambda$1502 and \SiII\,$\lambda$1485 features, respectively.
Only a lower limit on the age based on the fit to the \SiII\,$\lambda$1485 feature is obtained due to the relatively longer lifetime of late B-type stars.
Additional bursts cannot be entirely ruled out, but they are not necessary to reproduce the stellar features on the UV spectrum.
In contrast, the model spectra of continuous star-formation history provide poor matches to the line depths of both absorption features, as confirmed by the higher chi-square values.
At all ages the \SiII\,$\lambda$1485 is weaker than \SV\,$\lambda$1502.
The comparable strengths of \SiII\,$\lambda$1485 and \SV\,$\lambda$1502 highlight the bursty nature of the recent star formation in \src,
which is distinct from a continuous one observed in lower mass Lyman-break galaxies at comparable redshifts \citep{dMello2000,Pettini2000,Quider2009,Quider2010}.

The \textit{absolute} strengths of the \SiII\,$\lambda$1485 and \SV\,$\lambda$1502 photospheric lines provide additional evidence for a short star-formation timescale.
Both features are observed to be deeper than predicted by any model spectra explored here.
Given that both Si and S are $\alpha$-elements,
their over-abundance compared to the Sun is in accord with the expectation that \src\ would evolve to become a massive elliptical --- they are enhanced in $\alpha$-elements due to their short star-formation timescales \citep{Tinsley1979b, Greggio1983, Thomas1999, Thomas2005}.
The high abundance is further corroborated by the presence of photospheric metal lines such as \OIV, \OV, and \FeV\ as discussed in \S\ref{sec:photosphere_weak}.

The full comparison of the UV spectrum of \src\ to models is illustrated in Figures~\ref{fig:compare_sb99_instant} and \ref{fig:compare_sb99_const} for the respective star-formation histories. 
The model spectra only provide predictions of stellar features,
while it is apparent that additional features from AGN and the interstellar medium are certainly present in \src.
Below we make note of the anomalies in the comparison to the model spectra.
None of the model stellar spectra can reproduce the strong resonant emission lines (\NV, \SiIV, and \CIV) detected in \src. 
This implies that the AGN must be at least partially responsible for the emission of these resonant doublets.
As for the absorption components of the resonant lines, 
it is difficult to conclusively determine the age with these features for two reasons.
Firstly, absorption lines are partially filled by the emission components as well as nearby strong emission lines, 
e.g., \La\ is just blue-ward of \NV\,$\lambda$1238, and 
\SiII$^{\ast}$\,$\lambda$1533 is blue-ward of \CIV\,$\lambda$1548.
Secondly, stellar wind lines are stronger and have larger blueshifts in more metal-rich galaxies \citep{Leitherer2001}.
These complications may explain why none of the model spectra over the age range 1--50\,Myr can simultaneously reproduce the absorption components of all the resonance lines. 
Nonetheless, the \SiIV\ resonant doublet is well-reproduced by a young stellar population of a few Myr-old burst.
If the blue absorption edge of \SiIV\ is attributed to stellar wind,
evolved O-type supergiants are required to reproduce this feature.
They only appear between 3--6 Myr after an instantaneous burst \citep[see §\ref{sec:knots};][]{Conti1996}.

Some photospheric features that are expected to be in absorption are instead observed in emission in \src, as discussed in \S\ref{sec:photosphere_weak}.
Such features include the \CIII\,$\lambda\lambda$1426,1428 doublet and \CIII\,$\lambda$1620.
Moreover, it appears that the \NV\,$\lambda\lambda$1616,1621 doublet and \NIV\,$\lambda$1718 may have P-Cygni profiles. 
Wolf-Rayet stars are known to show emission or P-Cygni profiles in some lines that are normally in absorption in O-stars \citep{Willis1982, Kinney1993}.
The presence of Wolf-Rayet stars is suggested by the possible absorption components in \HeII\ as discussed in §\ref{sec:sysred} and §\ref{sec:photosphere}.
Unfortunately, the strongest Wolf-Rayet star features lie in the rest-frame optical, such as \NIII\,$\lambda$4640,
\HeII\,$\lambda$4686, and \CIV\,$\lambda$5808 \citep{Conti1991}.  
All these features are redshifted into the observed NIR, which is unfortunately plagued by sky emission lines and is not particularly sensitive.
Although we are unable to unambiguously determine whether Wolf-Rayet stars are present in \src, given the dominance of young stars in its UV continuum emission
and the young age of its UV-emitting stellar population, 
observing a contribution from Wolf-Rayet stars in the spectrum of \src\ would not be a surprise. 
On the other hand,
galactic winds have been suggested to be responsible for
the emission of excited fine-structure lines in distant star-forming galaxies (\SiII$^\ast$ and \FeII$^\ast$; \citealt{Shapley2003,Steidel2016,Finley2017}; but see also \citealt{DZavadsky2010}).
If these excited lines of different elements share a common origin,
their presence in galaxies of all star-formation histories, 
whether bursty or continuous,
might favour the interpretation that they originate from resonant scattering associated with galactic winds rather than from short-lived Wolf-Rayet stars.
We will explore the nature of these excited lines in a forthcoming paper.

We note that the blends of photospheric lines \SiIII\,+\,\CIII\ at 1294--1296\AA,
as well as \CII\,+\,\NIII\ at 1324\AA, are poorly reproduced in
\texttt{Starburst99} models.  A similar remark was made in the analysis of
the composite UV spectrum of lensed star-forming galaxies at $z\sim$2 \citep{Rigby2017}.

\subsubsection{Comparison with local starbursts: Starburst in knots} \label{sec:knots}

Only few nearby starburst galaxies have UV spectra of sufficiently high resolution to make for a useful comparison with our spectrum of \src.
In Figure~\ref{fig:spec_ngc} we illustrate the comparison to nearby starburst galaxies with photospheric features detected and have sufficiently high spectral resolution ($<0.5$\,\AA), i.e., observed with the Goddard High Resolution Spectrograph aboard the \textit{HST} \citep[GHRS; ][]{Brandt1994}\footnote{The spectra of nearby starbursting dwarf galaxies were retrieved from http://www.stsci.edu/science/starburst99/docs/templ.html}.
These starburst galaxies are all dwarf galaxies,
where the spectra cover only the central starburst regions due to the limited field-of-view of the instrument.
NGC\,4214 is a barred irregular galaxy,
while NGC\,1741 is a merger with two clusters in the centres of the merging galaxies.
The observed starburst knots in these two galaxies,  NGC\,4214-1 and NGC\,1741B1, contain Wolf-Rayet stars and have best-fitting ages of 4--5\,Myr \citep{Conti1996, Leitherer1996}.
On the other hand,
the super star cluster NGC\,1705-1 has a slightly older age, $\sim$10\,Myr \citep{Heckman1997b, dMello2000}.

The stellar features in the nearby starbursting dwarf galaxies are overall similar to those in \src\ with small differences, as seen in Figure~\ref{fig:spec_ngc}.
The \SiIV\ absorption wing of \src\ resembles that of the younger NGC 4214-1 and NGC 1741B1 rather than the slightly older NGC 1705-1.
This is in line with our findings presented in \S\ref{sec:sb99} that \src\ hosts a young stellar population of around 4--7\,Myr old.
Another noticeable difference between \src\ and nearby starbursting dwarf galaxies is the strength of the metal lines (e.g., \FeV, \OIV, \OV, \SV).
These metal absorption lines are prominent features in \src\ as discussed in \S\ref{sec:sb99},
but are weak or absent in nearby starbursting dwarf galaxies.
This is in accord with the expectation that the star formation in \src\ likely took place in gas pre-enriched by previous episodes of star formation that assembled the bulk of its stellar mass ($>$10$^{11}$\,\Msun).
As stellar mass is observed to be correlated with the metallicity \citep{Tremonti2004},
the absence of metal lines in nearby dwarf galaxies, having only 1/4 -- 1/2 solar metallicity, are as expected.

A remarkable observation is that the photospheric lines (e.g., \CIII\,$\lambda$1247, \SV\,$\lambda$1502) in \src\ are as narrow as those in the starburst knots as shown in Figure~\ref{fig:spec_ngc},
despite a few orders of magnitude difference in mass.
The intrinsic velocity dispersion of photospheric lines are $\sigma_{\star}\sim$ 0.6--3.0\,\AA~in the rest frame after correcting for velocity smearing by the instrument,
as listed in Table~\ref{table:photospheric}.
These line widths correspond to about $\sigma_{\star}\leqslant55$\,\kms, 
indeed comparable to those of the nearby starburst knots and in fact not much higher than those of Galactic giant \HII\ regions \citep[up to $\sim25$\kms; e.g.,][]{Melnick1987}.
The photospheric features have remarkably narrow dispersions when compared to similarly massive quiescent galaxies at $z\sim2$ \citep[$\sigma_{\star}\gtrapprox 250$\kms; e.g.,][]{Toft2012}.
The small stellar dispersions suggest that the recent star formation in \src\ is concentrated in starburst knots on a smaller scale than the host galaxy, and are kinematically decoupled from the gravitational potential.

Multiple factors contribute to broadening the photospheric lines observed in \src.
In addition to the bulk motion of stars,
photospheric lines are also broadened by processes intrinsic to stellar atmospheres, such as thermal and pressure broadening, as well as rotation \citep[see review in][Ch. 11]{Gray2008}.
Disentangling the effects of these processes is clearly beyond the scope of this paper.
Here, we take an educated guess of the most important processes that broaden the photospheric lines in \src.
For the OB-type stars that dominate the UV spectrum of \src, the atmospheres of these individual stars can have intrinsic velocities of tens or up to a couple of hundreds of \kms\ \citep{Slettebak1956, Stoeckley1968}.
Thermal broadening is likely more significant than pressure broadening to explain the line width of these hot stars, given the low density of the atmospheres of supergiants that dominate the UV spectrum of \src\ (see §\ref{sec:sb99}).
On the other hand, ALMA \CI\ observations of the cold star-forming gas of \src\ has narrow velocity dispersion of $\sigma_{\mathrm{[C \textsc{i}]}}\lesssim30$\,\kms,
as we elaborated in \citet{Lelli2018} and shall further discuss in \S\ref{sec:CI}.
These facts together suggest that the widths of the photospheric lines of \src\ are dominated by the stellar atmosphere rather than their kinematics within the host galaxy.
If true, the narrow intrinsic stellar dispersion may indicate gas compression in \src,
as we shall discuss in detail in \S\ref{sec:compression}.

\subsubsection{Multiple velocity components of star formation}\label{sec:vel_comp}

\begin{figure}[htbp]
\begin{centering}
\includegraphics[width=\linewidth]{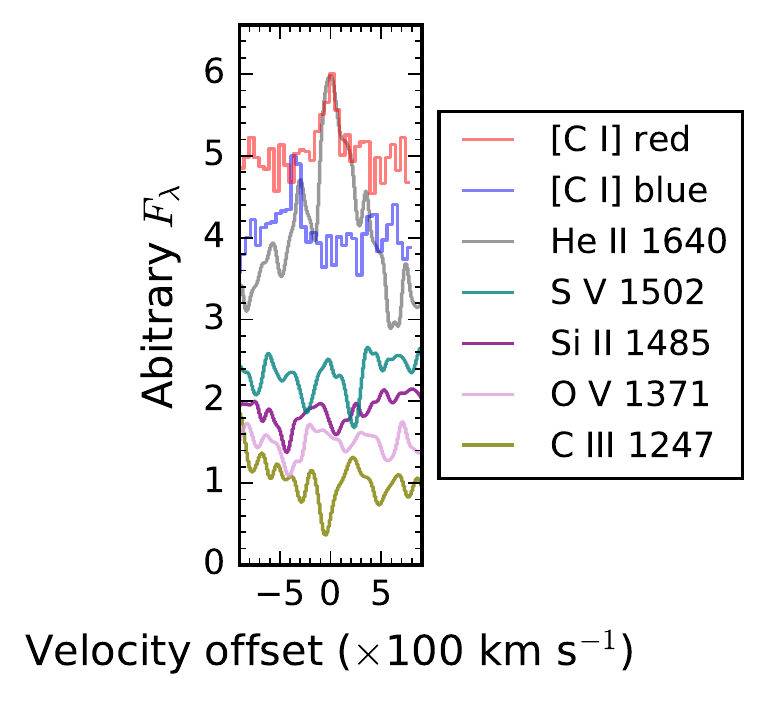}
\caption{
A comparison of the profiles of photospheric absorption lines (\CIII\,$\lambda$1247, \OV\,$\lambda$1371, \SiII\,$\lambda$1485, and \SV\,$\lambda$1502) to \HeII\,$\lambda$1640 and \CItwoone\ emission lines. 
\src\ has an asymmetric \HeII\,$\lambda$1640 emission line profile, 
consisting of two possible absorbers over a broad emission line and/or multiple narrower emission line components. 
This is in stark contrast with other high-redshift radio galaxies whose \HeII\ emission is dominated by broad emission. 
Note that the \HeII\ absorption at $\sim300$\kms\ is affected by faint sky lines and thus is not considered to be a real feature here.
The red and blue lines are the \CI\ spectra extracted from two separate regions from
the ALMA cube (Table~\ref{table:CI}). 
Note that the photospheric lines might consist of multiple velocity components corresponding to the \HeII\ emission or absorption lines, or they might be resolved into multiplets.
The \xs\ spectrum has been normalized to their continuum, Gaussian smoothed, and arbitrarily shifted upwards for easier visualization.
}
\label{fig:vel_comp}
\end{centering}
\end{figure}

\xs\ and ALMA have sufficient spectral resolution to resolve the profiles of the emission and absorption lines in \src.
In order to better characterize the nature of the peculiar \HeII\ emission line profile as discussed in \S\ref{sec:sysred} and \ref{sec:photosphere},
we compare it with the \CI\ emission as well as the photospheric lines that are not affected by nearby absorption features.
The two narrow \HeII\,$\lambda$1640 peaks coincide with the two ALMA \CItwoone\ velocity components.
The photospheric lines may also have at least two velocity components\footnote{We note, however, the challenge of determining velocity components based on photospheric lines alone,
given that all the photospheric lines plotted in Figure~\ref{fig:vel_comp} could be multiplets based on atomic line calculations (Footnote~\ref{footnote:atomicline}).
Without a reference high-resolution UV spectrum of a nearby galaxy or knowledge of their physical conditions,
%we lack information to deduce the relative strengths of these multiplets.
we are unable to distinguish between multiple velocity components or multiplets.
}.
Most notably, \CIII\,$\lambda$1247, \OV\,$\lambda$1371, and \SiII\,$\lambda$1485 have two absorption components that coincide with the emission of \HeII\ and \CI.
The three absorption components of \SV\,$\lambda$1502, on the other hand, coincide with the absorption components of \HeII. 
These observations may suggest that part of both the \HeII\,$\lambda$1640 emission and absorption could be due to young stars in addition to being photoionized by the AGN.
Attributing the emission and absorption at least partially to young stars could alleviate the tension between the observed \CIIIsf/\HeII\ and \La/\HeII\ ratios of \src\ compared to photoionization model predictions and other high-redshift radio galaxies \citep{Humphrey2008a}.
Deep, high-resolution spectroscopic observations in the UV and submillimeter are required to determine that the two velocity components in the UV correspond to two spatial components as suggested by the current data.
Overall, multiple velocity components are consistent with our finding that \src\ has undergone multiple bursts of star formation of different ages as discussed in \S\ref{sec:sb99}.

\subsubsection{Obscuration of the young stars in \src}\label{subsubsec:UVcont}

The slope and the polarization of the UV continuum informs us
whether the UV photons originate from massive stars or the AGN.
Spectropolarmetric observations of \src\ have placed a stringent upper-limit on the polarization 
of $<$12.3\% (from a deep VLT/FORS spectrum taken by JDRV; D. Buchard 2008 M.Sc. thesis).
Therefore, the lack of polarization implies that, unlike other radio galaxies \citep{Cimatti1993,Cimatti1997,Cimatti1998, Dey1996, Vernet2001}, 
the UV continuum of \src\ is dominated by starlight as in 4C\,41.17 \citep{Dey1997} and that its slope can be used to estimate the unobscured star-formation rate.  

By fitting the continuum with a power law while excluding regions with strong emission and
absorption lines, we estimate the best-fitting continuum between
rest-frame 1245--1534\,\AA~(UVB arm, roughly between \La\ and \CIV) to be
$f_{\lambda}=(9\pm3)\times10^{-18}$\,$\lambda_{\mathrm{obs}}$$^{-0.27\pm0.05}$,
where $f_{\lambda}$, the observed flux density, is in units of erg\,s$^{-1}$\,cm$^{-2}$\,\AA$^{-1}$,
and $\lambda_{\mathrm{obs}}$ is the observed wavelength in \AA.
On the other hand, the continuum between rest-frame 1668 and 2300
\AA\ (VIS arm, roughly between \OIIIsf\ and \OIII) is best-fitted with
$f_{\lambda}=(1.3\pm0.2)\times$10$^{-16}$\,$\lambda_{\mathrm{obs}}^{-0.61\pm0.02}$.
The UV slope\footnote{As the ADC in VLT/\xs\ was
disabled during the observations, slit losses could potentially affect
the observed continuum slope in the UVB and VIS arms despite observing
close to the parallactic angle for all observations. Reassuringly, a comparison
of our spectrum with a shallower, lower resolution NTT spectrum
\citep{Broderick2007} shows that the slopes are consistent and suggests that slit losses do not severely affect our measurement of the rest-frame UV continuum slope.}, \betaUV$=-0.27\pm0.05$, is redder than other HzRGs which have been observed
\citep{Vernet2001}.

We make use of the spectral constraints of the stellar age to break the dust-age degeneracy. 
The rest-frame 1500\,\AA\ luminosity is given by
\LUV = $\lambda f_{\lambda} \times 4\pi\dlum^{2}=(7.1\pm2.8)\times10^{43}$\,\ergs = $(1.8\pm0.7)\times10^{10}$\,\Lsun.
Using the best-fit power law to the continuum, 
we estimate that the specific luminosity at rest-frame 1500\,\AA\ to be 4.7$\times$10$^{40}$\,\ergs\,\AA$^{-1}$.
This specific luminosity corresponds to
an unobscured SFR of $\sim$40, 90, and 1590\,\Myr\ for a stellar population with ages 
of 4, 6, and 44\,Myr-old for a single burst, 
or 6--8\,\Myr\ for continuous star formation at the same age interval \citep{Leitherer1999}. 
The high star-formation rate and high infrared luminosity estimated by \citealt{Falkendal2018} (see Table~\ref{table:prop}), 1\,020~\Myr, is highly discrepant with our estimate from the UV continuum,
implying that the UV continuum must be heavily obscured.

So what fraction of star formation is obscured?
First, we use \texttt{Starburst99} model spectra for the relationship between the UV continuum slope and the age of a stellar population.  
For both instantaneous burst and continuous star-formation history, the intrinsic
\betaUV\ is $\sim-$2.5 for all ages $\la$10\,Myr, 
and is about $-1.5$ for a single $\sim$45\,Myr burst for our explored models \citep{Leitherer1999}.
Comparing these values to the observed \betaUV$\sim-$0.3, we infer an extinction of 1.2 -- 2.2\, mag.
This extinction estimate is much lower than that of the 
dust torus obscuring the AGN in \src\ \citep[A$_{V}$=34.1;][]{Drouart2012} and
consistent with that estimated for the 
narrow-line region \citep[A$_{V}$=1.6$\pm$1.0;][]{Humphrey2008a}. 
This supports our interpretation that the UV continuum traces region far beyond the AGN proper and is embedded within the narrow-line region on galaxy-wide scale.

To understand the implication of the dust attenuation, we compare \src\
to the empirical IRX-$\beta$ relation \citep{Meurer1999}.
The infrared excess IRX is defined as \LIR/L$_\textrm{1500\,\AA}$.
The IRX of \src\ is $\sim$10$^{2.6}$, 
which is significantly above any standard dust attenuation conventionally considered for the Milky
Way and the Magellanic Clouds,
and falls well in the region of dusty star-forming galaxies at $z=2-3$ \citep{Nordon2013,Casey2014}.
Possible explanations for the boosted IRX/$\beta$ ratio compared to conventional dust attenuation laws include a bursty star formation history, high turbulence, low obscuration fraction of stellar light, a mixed
distribution of stars and dust (as opposed to a dust screen geometry),
and high optical depth in birth clouds surrounding young stars \citep{Calzetti2001,Xu2004,Boquien2009,Wild2011,Casey2014,Popping2017b}.
All these effects are plausible in \src,
and most certainly we have evidence for bursty star formation (\S\ref{sec:sb99}).
Determining the relative importance of these effects is beyond the scope of this paper.

While high \LIR/\LUV\ ratio is usually attributed to highly obscured star formation,
this inference assumes implicitly that star formation has been continuous,
such that the SFRs measured at UV and IR are sensitive to the same period of time.
In \S\ref{sec:sb99} we have shown that \src\ has undergone bursts of star formation in the past $\sim100$\,Myr.
An alternative and equally viable explanation for its discrepant IR and UV luminosities is a rapid truncation of star formation in the recent past.
UV continuum traces recent star formation while FIR is sensitive to longer timescales \citep{Hayward2014}.
This is because A-stars can continue to heat the dust and thereby power the FIR emission, while contributing little to the UV emission.
The implication of this interpretation will be further discussed in \S\ref{sec:discussion_gal}.

\subsection{Resemblance to 4C\,41.17} \label{subsec:4C4117}

The rest-frame UV spectrum of \src\ is strikingly similar to that of 4C\,41.17 (Figure~\ref{fig:spec_hzrg}), 
a high-redshift radio galaxy that has been widely cited as an example of AGN-triggered star formation \citep{Dey1997,Bicknell2000}.
The evidence for triggered star formation is the detection of a young stellar population through the photospheric absorption line \SV\,$\lambda$1502 and the blue absorption wing of \SiIV.
Both of these features are strongly detected in \src,
along with additional photospheric features thanks to the high signal-to-noise ratio and spectral resolution of the \xs\ spectrum.
In addition to the stellar UV absorption lines,
there are several other similarities between \src\ and 4C\,41.17.
Both UV continua show very low polarization and relatively flat slopes consistent with intense, heavily obscured star formation. 
The resonant lines, \NV, \SiIV\ and \CIV, have similar structure in both HzRGs.  
The two sources are comparably radio-luminous.
Although 4C\,41.17 is the only radio galaxy with both a high star-formation rate and a deep rest-frame UV spectrum to compare with,
this does show that it is possible for other high redshift radio galaxies to exhibit a spectrum as we have observed.
Summing up, these similarities suggest that \src\ is likely in a similar evolutionary stage of high-redshift radio galaxies as 4C\,41.17.
We will discuss the implications in a broader context in \S\ref{sec:discussion}.

\section{Result II: Low cold gas-mass fraction and high star-formation efficiency} \label{sec:CI}

\begin{figure}[h!]
\begin{centering}
\includegraphics[width=\linewidth]{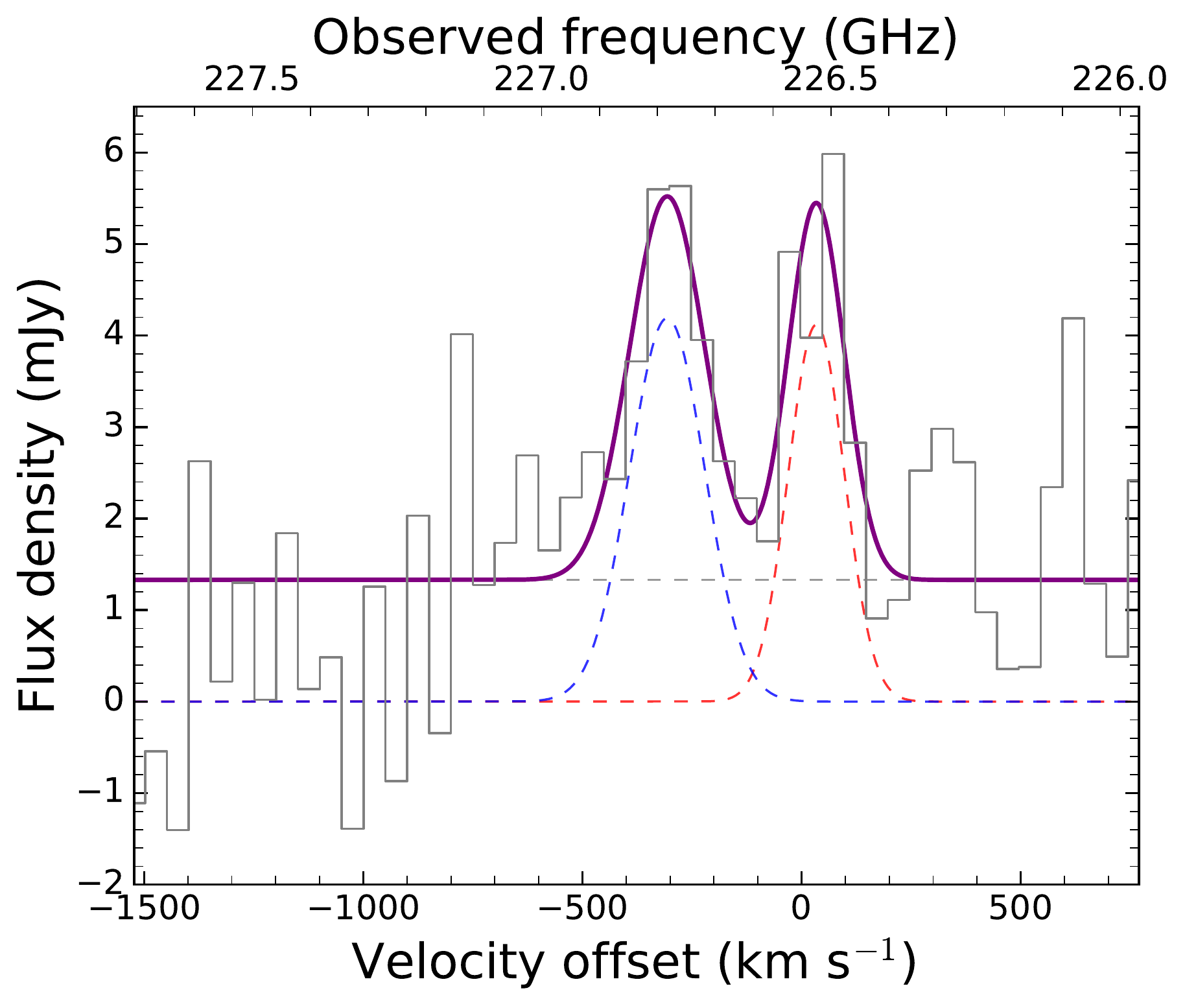}
\caption{The integrated \CItwoone\ spectrum of \src.
The best-fitting model (solid purple) is superposed on the Band 6 ALMA spectrum (grey histogram).
The individual components are plotted as dashed lines,
including the two independent Gaussian components (blue and red) and the continuum (grey horizontal).
\label{fig:CI_spec}}
\end{centering}
\end{figure}

\input{table_CI.tex}

The atomic carbon \CIlevels\ emission line at $\nu_\mathrm{rest}=809.34197$\,GHz, hereafter \CItwoone, is a tracer for the \Htwo\ gas mass \citep{Weiss2003,Papadopoulos2004a,Walter2011}.
\CItwoone\ has a critical density of ($1.2\times10^{3})$\,cm$^{-3}$, comparable to CO J=1--0 and probes diffuse molecular gas.
\CI\ is thought to be a better tracer of \Htwo\ mass than CO,
because it is less sensitive to variations in metallicity and evolutionary stage of the giant molecular clouds \citep{Glover2016}.
In particular \CI\ is argued to be a preferred tracer of molecular \Htwo\ in vigorously star-forming galaxies like \src,
in which the lower J level CO lines may become optically thick, and the cosmic ray intensity in starburst galaxies or AGN hosts may be high enough to dissociate CO molecules \citep{Papadopoulos2004a, Bisbas2015, Bisbas2017}.

In Figure~\ref{fig:CI_spec} we show the ALMA \CItwoone\ spectrum,
extracted using an elliptical aperture centered at $05^{h}30^{m}25^{s}.438 -54^{\circ}54'23''.135$,
with major and minor axes corresponding to $0\arcsec.64\times0\arcsec.43$ at a position angle of 20$^{\circ}$.
The extraction aperture was defined using the continuum-subtracted moment-0 map around the \CItwoone\ line.
The \CItwoone\ emission is clearly detected at the $H$-band continuum position and therefore encompassed by the \xs\ slit (Fig.~\ref{fig:slit}).
The \CI\ emission is near, but not exactly co-spatial with the western radio lobe.
No \CItwoone\ emission is detected at the position of the eastern lobe.
While the CO (7-6) line is observed in the same spectral window,
it is too close to the edge of the bandpass to be useful.
The \CItwoone\ emission on the western lobe is double-horned,
and the two velocity peaks are separated by $\sim0\arcsec.45$ or $\sim4$\,kpc in projection on the sky.
We have already presented a detailed dynamical model of the \CItwoone\ emission in \citet{Lelli2018}.
The \CItwoone\ emission is well-described by a dynamically cold rotating disk with rotation velocity of $\sim310$\,\kms\ and intrinsic dispersion of $\sigma_{\mathrm{[C \sc{I}]}}\lesssim30$\,\kms (see \S\ref{sec:knots}).

To measure the \CItwoone\ line flux for \src,
we fit the extracted 1D spectrum with two Gaussian components and a continuum as shown in Figure~\ref{fig:CI_spec}.
The ALMA Band 6 continuum of the host galaxy is $(1.33\pm0.16)$\,mJy \citep[][Table~A.13 western component]{Falkendal2018},
which was estimated using the full bandwidth of four spectral windows with line channels excluded.
Given the narrow bandwidth covered by each ALMA spectral window,
the continuum can be approximated as a flat, straight line.
The best-fitting parameters of the two Gaussian components are listed in Table~\ref{table:CI}.
Integrating the line emission, 
we find a total velocity-integrated flux $S\Delta v=(2.0\pm0.5)$\,Jy\,\kms.

We first estimate the neutral atomic carbon mass from the \CItwoone\ line luminosity,
following the approach derived in \citet{Papadopoulos2004a}.
We assume local thermodynamic equilibrium, that the \CI\ lines are optically thin, and that the excitation factor $Q_{21}$=0.5.
The excitation factor $Q_{21}$ is 
defined as the fraction of the column density originating from the upper level transition, 
is a factor ranging from 0 to 1.
$Q_{21}$ is a function of the temperature, density and radiation field,
none of which we can directly constrain without observations of other lines such as \CIonezero\ or multiple CO lines.
We note, however, that the uncertainty in $Q_{21}$ can only affect the gas mass estimate by a factor of two at most, comparable to measurement errors.

To convert the neutral carbon mass to global molecular hydrogen mass \MHtwo, we adopt an \CI-to-\Htwo\ abundance,
\XCI~=~3$\times$10$^{-5}$ following the value derived for M82 \citep{Weiss2003} and adopted in subsequent studies \citep{Papadopoulos2004b,Wagg2006,Gullberg2016b}. 
The conversion to molecular mass includes the mass in Helium by a correction factor of 1.36 \citep[e.g.,][]{Solomon2005}.
The adopted value of \XCI\ agrees well with that predicted for metal-rich galaxies like \src\ \citep{Glover2016}.
The systematic uncertainties to measuring the \Htwo\ gas mass with \CI\ are discussed in detail in \citet{Papadopoulos2004a}, 
and appear comparable to the measurement uncertainty in our case. Noting the linear dependence of \MHtwo\
on \XCI\ and $Q_{21}$, 
we estimate based on theoretical studies \citep{Glover2015,Glover2016} that the \CI-to-\Htwo\ conversion has a systematic uncertainty of at most $\sim3$.
Indeed, \Htwo\ masses inferred from \CI\ and CO agree within a factor of 2--3 for dusty star-forming galaxies with properties similar to \src\ \citep{AZadeh2013,Bothwell2017}.
The derived \MHtwo\ values are listed in Table~\ref{table:CI} for the two velocity components, 
amounting to a total molecular mass of \MHtwo~$=(3.9\pm1.0)\times10^{10}$\,\Msun.
Its molecular gas fraction, defined as \fgas $\equiv$ \MHtwo/(\MHtwo+\Mstar), is only $(12\pm7)$\%.
This implies that \src\ is nearing the end of its star formation and will soon run out of fuel\footnote{The depletion of cold ($T\lesssim50$\,K), star-forming gas does not imply the lack of warm ($T\sim10^{4}$\,K) gas. On the contrary, giant \La\ reservoir is a common feature of HzRGs including in \src\ (\S\ref{sec:sysred}).} and quench.

We estimate the star-formation efficiency, SFE~$\equiv$~SFR/\MHtwo, of
the entire system. 
Using \MHtwo\ derived here and the SFR, 
we find SFE~$=(26\pm8)$\,Gyr$^{-1}$.
The depletion timescale, defined as the inverse of SFE, is then \tdepl~$\equiv$~SFE$^{-1}$~$=(38\pm12)$\,Myr.
We provide these values in Table~\ref{table:hzrg_gas}.
These results imply that \src\ is forming stars more efficiently than the Milky Way by at least an order-of-magnitude.
In \S\ref{sec:ks} we further derive the relation between the surface densities of gas and star formation in \src, 
and discuss it in context with other galaxies at comparable redshifts and stellar masses.

\section{Discussion}\label{sec:discussion}

Our multi-wavelength analysis makes \src\ one of the high redshift galaxies with best constraints of star-formation history and conditions.
What can we learn about AGN feedback and star formation quenching from this detailed investigation?
What is the relation between its radio jets and extreme star formation
--- a mere coincidence, a co-evolution, or a causality?
In this Section, 
we address these questions by making use of the exquisite insights from our analysis of \src\ and compare its properties with various galaxy populations.

\subsection{A synthetic picture: quenching via gas depletion}\label{sec:gas_depl}

\begin{figure}[!ht]
\begin{centering}
{
\includegraphics[width=\linewidth]{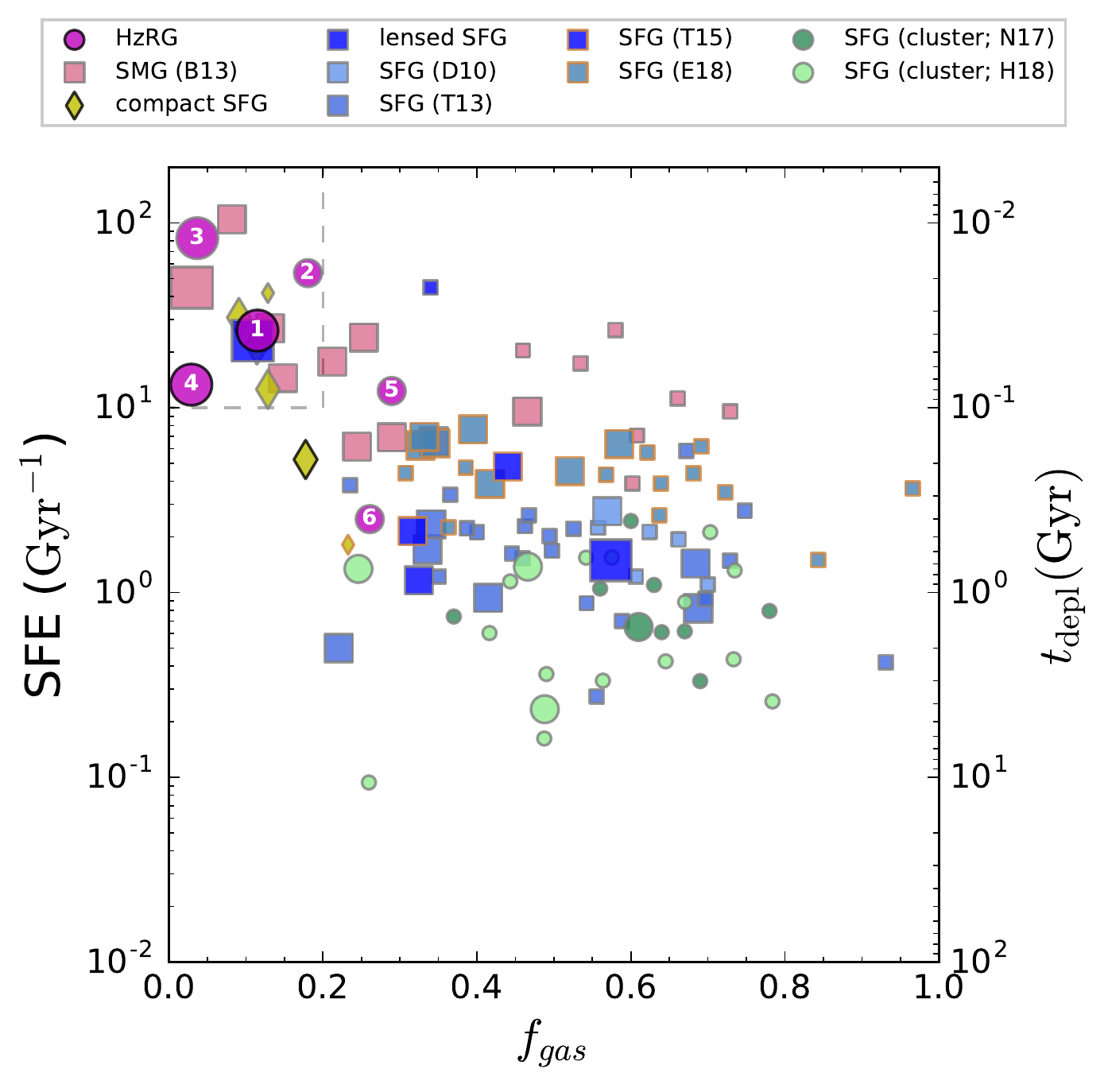}}
\caption{
Star-formation efficiency (SFE $\equiv$ SFR / \MHtwo) as a function of molecular gas fraction \fgas $\equiv$ \MHtwo/(\MHtwo+\Mstar) of \src\ and other galaxies at $z=1.3-4.1$.
High-redshift radio galaxies are plotted as magenta circles, 
individually labeled as (1) \src, the subject of this work; (2) 4C\,41.17; (3) MRC\,0152-209, the Dragonfly; (4) MRC\,1138-262, the Spiderweb; (5) 4C\,60.07; and (6) MRC\,0943-242.
Submillimeter galaxies are plotted as pink squares.
Star-forming galaxies are plotted as blue squares, 
with different shades referring to the respective samples.
Compact star-forming galaxies are plotted as yellow diamonds.
Cluster galaxies are plotted as green circles.
Full references are provided in \S\ref{sec:gas_depl}.
The symbol size denotes the stellar mass of each galaxy split in three bins of \Mstar $\geqslant3\times10^{11}$\Msun, $1-3\times10^{11}$\Msun,
and $<10^{11}$\Msun, where larger symbols represent more massive galaxies.
The edge colours of each symbol indicate the \MHtwo\ tracer:
black for \CI, grey for CO, and brown for dust continuum.
The grey dashed lines mark the corner for the efficiently star-forming and gas-poor galaxies.
\label{fig:sfe}
}
\end{centering}
\end{figure}

The relationship between the radio jets and star formation, if any,
can be investigated through the star-formation efficiency, 
the star-formation rate per unit cold gas mass.  
To understand the implications of the SFE derived in \S\ref{sec:CI},
in Figure~\ref{fig:sfe} we plot the SFE and \tdepl\ as a function of \fgas together with several galaxy populations at $z=1.3-4.1$ with SFR, \Mstar, and \MHtwo\ measurements through \CI, CO, or dust continuum.
Whenever relevant, the masses have been adjusted to the cosmology and IMF used in this work. 
We also estimate the SFE for other radio galaxies with high SFRs and compact radio sources. 
The high-redshift radio galaxy (HzRG) sample includes \src\ \citep[this work; ][]{Falkendal2018},
4C\,41.17 \citep{Dey1997,dBreuck2005,dBreuck2010,Drouart2016}, 
the Dragonfly galaxy \citep[MRC\,0152-209;][]{dBreuck2010,Emonts2015a,Falkendal2018},
the Spiderweb galaxy \citep[MRC\,1138-262;][]{Hatch2009,Gullberg2016b},
4C\,60.07\ \citep{Greve2004,dBreuck2010},
and MRC\,0943-242 \citep{Gullberg2016a}.
In the case of the Dragonfly and MRC\,0943-242, only the host galaxies are considered rather than the extended gas reservoir and nearby companions.
All six HzRGs have \MHtwo\ estimated from either the \CI\ or low J-level CO emission lines.

We also overplot a compilation of star-forming galaxies at $z=1.3-4$ with \MHtwo\ measurements.
These include the colour-selected sample\footnote{Their \MHtwo\ have been adjusted to a Milky Way CO-to-\Htwo\ conversion factor to match the conversion factor of \citet{Tacconi2013}.} of \citet{Daddi2010} and the PHIBBS sample \citep{Tacconi2013} which are both CO surveys,
the ALMA dust continuum samples of SFGs at $z=1.3-3.2$ \citep{Elbaz2018} and at $z=2.53$ \citep{Tadaki2015},
as well as several lensed SFGs with dust continuum and CO detections \citep{Bothwell2013b,Sharon2013,DZavadsky2015,Nayyeri2017}.
Compact SFGs \citep{Tadaki2015,Tadaki2017,Spilker2016,Barro2017,Popping2017a} and 
luminous submillimeter galaxies (SMGs) from the sample of \citet{Bothwell2013b} are also plotted.
To investigate potential effects due to environment,
we also plot two cluster galaxy samples at $z=1.5-1.6$ \citep{Noble2017,Hayashi2018}.

Several observations can be made from Figure~\ref{fig:sfe}.
The HzRGs plotted here in general have lower \fgas than most star-forming galaxies at high redshifts.
The first four HzRGs including \src\ have \fgas$\la$ 20\% that puts them on par with local spiral galaxies
\citep[e.g.,][]{Leroy2008}.
Considering the broader population of HzRGs including those that do not have gas mass measurements and therefore not shown in Figure~\ref{fig:sfe},
HzRGs overall likely have even lower \fgas as most of them are undetected in \CI\ searches (T. Falkendal, private communication).
Since HzRGs are the most massive galaxies at $z=0-4$ \citep{Miley2008},
their low \fgas is as expected if they would soon quench their star formation.
Another observation is that HzRGs plotted in Figure~\ref{fig:sfe} appear more efficiently star-forming than normal disk galaxies at $z\sim2$ by an order-of-magnitude.
The difference is larger than any systematic uncertainty inherent to estimates of \Htwo\ mass (such as abundance, excitation correction) or SFR (such as the IMF, dust correction, calibration).
Lastly, we find no evidence for the environment in driving the star formation activity and gas depletion,
other than perhaps enhancing the merger rate,
given that HzRGs reside in dense environments and are relatively gas depleted compared to the brightest, ALMA-detected cluster galaxies.

Restricting our comparison to massive galaxies (\Mstar~$>10^{11}$\Msun), 
high SFE ($>10$\,Gyr$^{-1}$), 
and low gas fractions (\fgas$\leqslant20$\%),
we identify other classes of galaxies which are as depleted in gas and yet as efficiently star-forming.
A subset of SMGs match these criteria (\citealt{Bothwell2013a}; see also \citealt{Ivison2011}).
including the lensed dusty star-forming galaxy, NA.v1.489, which is also a SMG \citep{Nayyeri2017},
as well as most compact star-forming galaxies.
While it can be argued that their higher SFE and lower \fgas might be attributed to the choice of CO-to-\Htwo\ conversion factor that could vary by a factor of $\sim4$,
we highlight that the discrepancy still stands if we only consider the galaxies with \CI\ measurements and thus unaffected by this systematic uncertainty.
What all these galaxies have in common is their high cold gas surface densities,
with extreme star formation confined to only a few kpc,
likely driven by compressive gas flows and/or mergers,
as we shall discuss in \S\ref{sec:ks}.
In fact, several of these galaxies are shown to be ongoing mergers or merger remnants, including the Dragonfly galaxy \citep{Emonts2015b,Emonts2015a}, the Spiderweb galaxy \citep{Gullberg2016b}, 4C\,41.17 \citep{dBreuck2005}, 
and one of the four SMGs in this region \citep[SMMJ123711+622212; ][]{Bothwell2013a}.
In \S\ref{sec:trigger} we discuss the possibility of \src\ as an interacting system.
This merger fraction of $\sim$4/12 is significantly higher than comparably massive galaxies at $z\sim2$ \citep{Man2016b,Silva2018}.
In addition to the HzRGs, there is at least another galaxy in that region that hosts an AGN \citep[SMMJ030227+000653; ][]{Bothwell2013a}.
Therefore the AGN fraction is also high, at least 5/12.
Our findings corroborate several studies reporting low \fgas\ and short \tdepl\ among $z>1$ AGN host galaxies compared to matched non-AGN samples \citep{Brusa2015,Fiore2017,Kakkad2017}.
While the data at hand are insufficient for us to infer whether AGN or mergers are responsible for compressive gas motions,
their prevalence among galaxies with high SFE and low \fgas suggests that both are plausible mechanisms.
We will discuss possible triggers of gas compression in \S\ref{sec:trigger}.

\input{table_hzrg.tex}

\subsection{Gas compression as the key to efficient star formation}\label{sec:compression}

\subsubsection{Clues from observations}\label{sec:ks}
\begin{figure}[!ht]
\begin{centering}
\includegraphics[width=\linewidth]{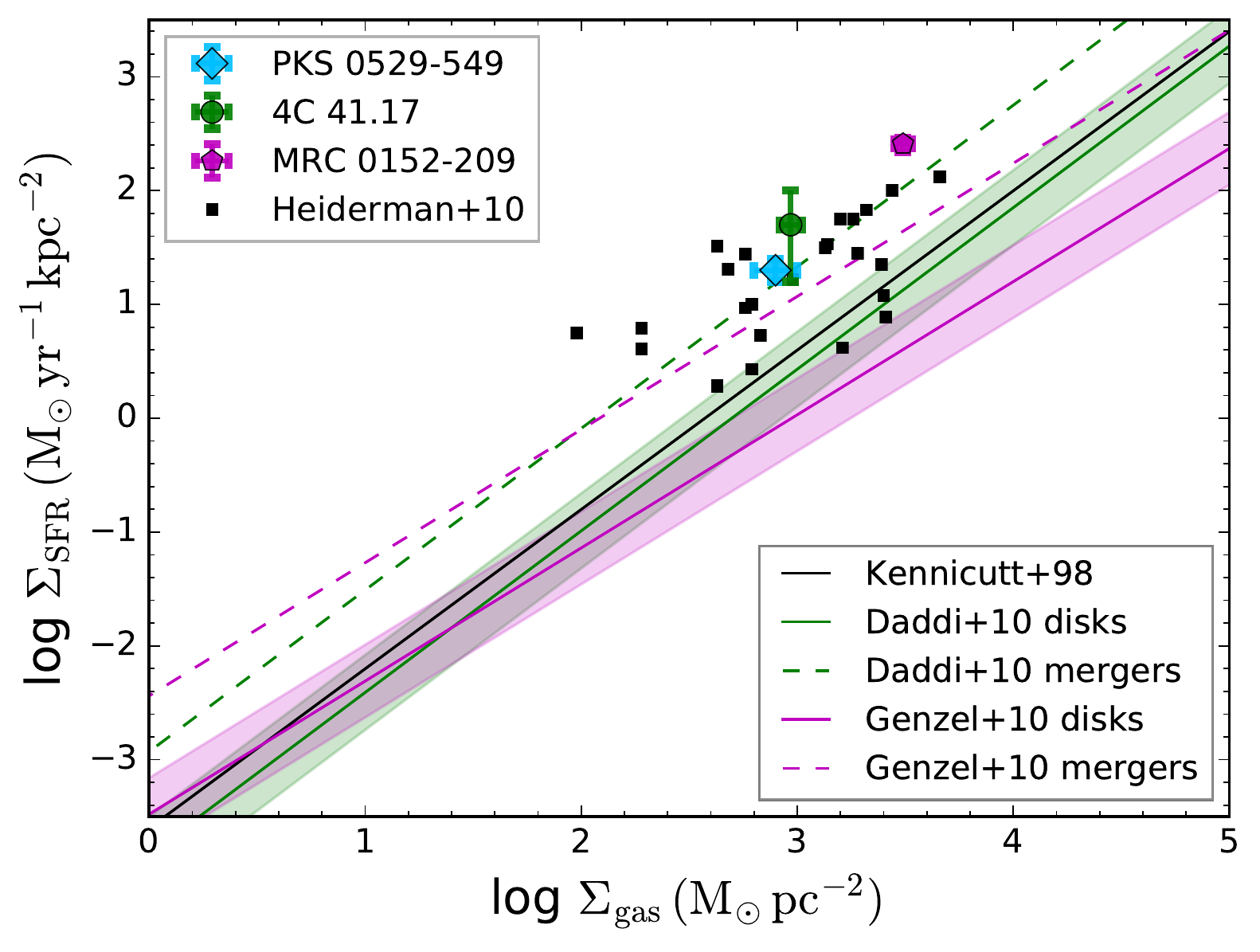}
\caption{
The surface density of gas, $\Sigma_\mathrm{gas}$, versus the
surface density of the star-formation rate, $\Sigma_\mathrm{SFR}$ \citep[the "Schmidt-Kennicutt relation";][]{Kennicutt1998b}.
We show the locations of three high-redshift radio galaxies:
\src\ (blue diamond; this work);
4C\,41.17 (green circle),
and MRC\,0152-209 (magenta pentagon).
The black squares denote the Galactic massive HCN clumps from \citet{Heiderman2010}.
The lines indicate the best-fit to the data presented in the respective works,
including local star-forming galaxies and circum-nuclear starbursts \citep[black;][]{Kennicutt1998b}, 
gas-rich galaxies at redshifts $\sim$1-3 \citep[magenta;][]{Genzel2010},
and colour-selected star-forming galaxies at $z\sim$1.5 \citep[green;][]{Daddi2010}.
Solid and dashed lines refer to the best-fit for disk star-forming galaxies and mergers, respectively.
}
\label{fig:SK}
\end{centering}
\end{figure}

Aside from star-formation efficiency arguments, 
we can discern the role of jets in influencing star formation through the Schmidt-Kennicutt relation \citep{Kennicutt1998b}.
If jets can induce star formation, 
one might expect the surface density of star formation, $\Sigma_\mathrm{SFR}$, 
to be enhanced relative to the surface density of the molecular gas, $\Sigma_\mathrm{gas}$.
We estimate $\Sigma_\mathrm{SFR}$ and $\Sigma_\mathrm{gas}$ for \src\
using the SFR estimate from Table~\ref{table:hzrg_gas}, 
gas mass inferred from the \CI\ emission (\S\ref{sec:CI}), 
and use the deconvolved radius of the \CI\ disk as the size of the star-forming region \citep[$R\sim4$\,kpc; ][]{Lelli2018}.
As in Figure~\ref{fig:sfe},
we also estimate $\Sigma_\mathrm{SFR}$ and $\Sigma_\mathrm{gas}$ for
4C\,41.17 and the Dragonfly galaxy as listed on Table~\ref{table:hzrg_gas}. 
For the Dragonfly galaxy we estimate $R$ to be 1.5\,kpc based on high-resolution CO (6-5) imaging \citep{Emonts2015b}.
For 4C\,41.17, we estimate $R$ to be half of the extent of the optical emission \citep[][adjusted to the same cosmology]{Miley1992}.
Although the Spiderweb galaxy is comparable in terms of SFE and \fgas\ to these three HzRGs,
there are only upper limits on both the SFR and the size \citep{Gullberg2016b} and therefore no meaningful constraint on $\Sigma_\mathrm{SFR}$ can be obtained with present data.
We compare these results to other studies of the Schmidt-Kennicutt relations presented in the literature \citep{Kennicutt1998b,Daddi2010,Genzel2010}. 

Our sample of high redshift radio galaxy hosts lie above the standard relationship as shown in Figure~\ref{fig:SK}.
Their $\Sigma_\mathrm{SFR}$ is on par with the most efficiently star-forming regions within the Milky Way \citep[massive dense clumps and young stellar objects;][]{Wu2010,Heiderman2010}.
This suggests that star formation in these HzRGs is governed by a process that dominates the disk-averaged self-gravity picture outlined in \citet{Kennicutt1998b},
wherein disk-wide star formation is assumed to scale with the growth rate of perturbation in the gas disk.
As the free-fall timescale is inversely proportional to density,
the offset of the HzRGs from the Schmidt-Kennicutt relation lends further support to our interpretation that the efficient star formation in radio galaxies including \src\ is localized in regions of compressed gas (\S\ref{sec:knots}).

\subsubsection{Theory} \label{sec:compression_theory}
\begin{figure}[!ht]
\begin{centering}
\includegraphics[width=1.1\linewidth]{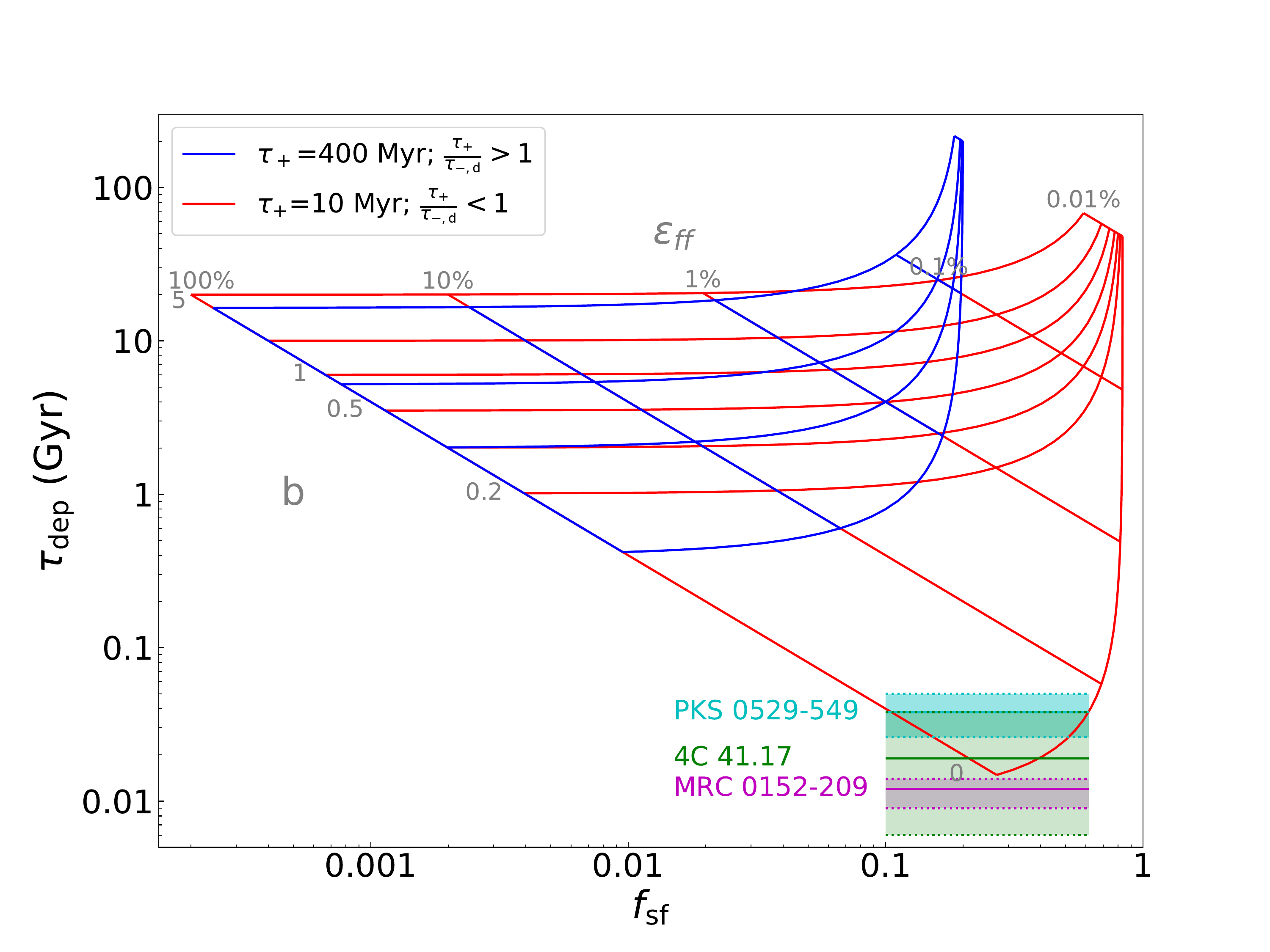}
\caption{
An analytic model for the relation between the star-forming fraction of the total gas mass (\fsf) and the gas depletion time (\tdepl) reproduced from \citet[][Figure 7]{Semenov2018}. 
The grids show the predicted \fsf\ and \tdepl\ for various values of the star-formation efficiency per free-fall time (\eff; how much of the gas is converted to stars on a cloud dynamical time), 
and the boost factor of momentum gain per supernova (b). We only show the b-values for the red grid lines.
The key parameters of this model are the timescale for gas to become star-forming ($\tau_{+}$) and the timescale for the star-forming gas to be rendered non-star forming ($\tau_{-,d}$).  
The blue grid is for local disk galaxies with long $\tau_{+}$ (400\,Myr) and the high ratio of $\tau_{+}$/$\tau_{-,d}$ (5). 
The red grid shows the case where $\tau_{+}$ is short (10\,Myr) and $\tau_{+}$/$\tau_{-,d}$ is low (0.1). 
It is apparent that local disk galaxy models are inconsistent with the short \tdepl\ ($\leqslant$100\,Myr) observed in the three high redshift radio galaxies including \src.
Gas compression can shorten \tdepl\ by reducing $\tau_{+}$ (therefore low $\tau_{+}/\tau_{-,d}$),
resulting in high \fsf, i.e., high dense molecular gas fraction.
}\label{fig:analyticmodel}
\end{centering}
\end{figure}

Strong gas compression and the concomitant rapid dissipation of mechanical energy of the gas flows would necessarily result in stellar systems with low velocity
dispersions if the young stars formed coherently.
This is regardless of whether the gas flows are driven by gravity (in case of mass transfer during a merger), or radio jets passing through the medium inducing instabilities, shocks or back flows. 
Our findings of \src\ support this scenario:
the young stellar population of \src\ has narrow intrinsic dispersion of only a few tens of \kms\ as deduced from the stellar photospheric absorption lines (\S\ref{sec:knots}),
and similarly for the cold, star-forming gas from the \CI\ spectrum (\S\ref{sec:CI} and \citealt{Lelli2018}).
These narrow lines suggest strong and rapid dissipation of the mechanical energy in the gas.

We can quantify this scenario of efficient star formation due to rapid compression and dissipation in the context of recent theoretical studies on the subject \citep{Semenov2017,Semenov2018,Elmegreen2018}.  
The analytic framework of \citet{Semenov2017,Semenov2018} is used here to guide our interpretation for the efficient star formation in \src\ and other radio galaxies. Their model uses simple relations between the gas depletion timescale and the mass-loading factor (the mass rate of return of the gas to a non-star-forming state relative to the star-formation rate), and three timescales:
(1) the time for gas to become star-forming ($\tau_{+}$);
(2) the free-fall time of the cloud ($\tau_{\mathrm{ff}}$);
and (3) the time for star-forming gas to become non-star-forming ($\tau_{-,d}$). 

In Fig.~\ref{fig:analyticmodel}, we show two grids based on this analytic model. 
The first model describes local star-forming disk galaxies: 
the timescale for gas to become star-forming is long ($\tau_{+}=$~400\,Myr, on the order of their disk orbital timescales), 
relative to the timescale to become non-star forming ($\tau_{-,d}\leqslant$100\,Myr). 
This results in long gas depletion timescales and a relatively low fraction of dense molecular gas,
in agreement with observations of nearby spiral galaxies \citep[e.g.,][]{Leroy2013, Usero2015}. 
The second model describes starburst galaxies such as \src\ that are undergoing strong compressional flows: 
the timescale for a molecular cloud to collapse remains the same, 
4\,Myr, but now, the timescale to become star-forming is very short, 10\,Myr.  
This follows previous studies suggesting that under high pressure the conversion
rate of gas to becoming molecular would be very high \citep[e.g.,][and many other studies]{Lehnert2009,Lehnert2013,Guillard2015}. 
The additional pressure and compression would also imply that the mechanical and radiative energy output from the massive stars would be injected into an overall denser, higher filling factor environment,
leading to less effective feedback.  
In summary, the two models vary mostly by $\tau_{+}$, with roughly similar $\tau_{-,d}$ values.
The short \tdepl\ of the three radio galaxies discussed in detail is inconsistent with the first model based on local star-forming disk galaxies.
Rather, these radio galaxies require the starburst model to explain their short \tdepl, under the following conditions:
the star-formation efficiency per free-fall time\footnote{Note that \eff\ is not equivalent to SFE. The former is defined using the free-fall time of a cloud in simulations, while the latter is an observational definition based on the molecular \Htwo\ mass.}, \eff, is high (between a few \% to 100\%);
the impact of feedback from massive stars appears weak, $0\leqslant b <<0.2$;
with a high fraction of star-forming gas, \fsf~$\ga$10\%  (i.e., high dense molecular gas fraction).

In summation,
star formation is made more efficient by compressional gas flows,
which are key to both changing the phase distribution of the ISM to be mostly molecular and high density and also, 
at least temporarily, quench the feedback from massive stars. 
The compression of the gas then results in stellar populations with modest velocity dispersions of tens of \kms\ as we observe in the young stars and cold gas of \src.
Thus this rules out the predictions of high-velocity motions in the clouds or stars often seen in theoretical models or simulations \citep[e.g.,][]{Zubovas2013a,Dugan2014}.

\subsubsection{Possible triggers for gas compression and depletion}\label{sec:trigger}

In previous subsections we have argued for the significant role of gas compression in increasing the efficiency of star formation. More specifically, we argued that the gas compression likely leads to a higher fraction of dense gas compared to normal star-forming galaxies
which then leads to a higher efficiency.
It is therefore interesting to identify mechanisms that might trigger galaxy-wide gas compression and gas depletion. 
Multiple processes are taking place simultaneously in \src,
including intense star formation, accretion onto the galactic nucleus and jets and ionized gas outflows associated with the supermassive blackhole,
and a possible merger. The multiplicity of processes makes it challenging
to attribute any galaxy-wide gas compression to a single process.  Here,
we discuss plausible mechanisms of gas compression and depletion, and
use timescale arguments to estimate their relative importance.

Radio jets have been suggested to compress the interstellar
medium to high densities by many theoretical studies
\citep{Fragile2004,Fragile2017,Silk2005,Silk2013,Gaibler2012,Dugan2014}.
This form of positive feedback effectively consumes the molecular
\Htwo\ in the galaxy through episodes of efficient star formation. 
This appears to explain the efficient star formation of the HzRGs plotted in Figure~\ref{fig:sfe},
all of which are compact radio sources.
If we use typical scaling of the speed at which hot spots advance
\citep{Carilli1991}, it is clear that the radio source of \src\ is young ($\la$\,Myr old). The most recent burst of star formation may be consistent with
the age of radio source, a few Myr, but not the much older burst of tens of
Myr old.  Observations of the warm ionized gas \OIII\ and \Ha\ have
already confirmed the existence of fast AGN-driven outflows in \src\
\citep{Nesvadba2017}, although we note that the warm ionized gas mass
is only $\sim6\%$ of the \Htwo\ mass as we derived in \S\ref{sec:CI}.
In addition, the outflow of warm ionized gas likely does not play a role
in triggering star formation in \src.  This is because the star-forming
ALMA \CI\ gas is misaligned with the ionization cone probed by \OIII\
and \Ha\ \citep{Lelli2018} and the narrow stellar dispersion is at odds
with the divergent gas outflows in AGN models. 
Overall, while the young radio jet may have triggered the most recent burst,
the AGN is unlikely to have triggered gas compression due to its divergent nature of gas flows.

This is not to say that the outflow has no impact on the star
formation. In addition to gas consumption through star formation, AGN outflows
may work in concert to clear out the remaining gas and make quenching
effective over a longer term.  The AGN and star formation in \src\ have
comparable bolometric luminosities ($L_{\mathrm{AGN}}=6\times10^{12}
$\,\Lsun\ and $L_{\mathrm{SF}}=9\times10^{12}$\,\Lsun;
\citealt{Falkendal2018}). Thus, if radiation pressure is effective in
driving outflows then this would imply that the AGN and massive stars appear to have roughly equal
contribution in dispersing the gas. The timescale estimated to disperse
the warm ionized gas \citep{Nesvadba2017} is much longer than the gas
depletion timescale suggesting that star formation plays a more important
role in consuming the gas than the outflow does in dispersing it.
And of course, this estimate only includes the mass of the warm ionized gas.
If we include the mass of the molecular gas, the timescale for the outflow to
drive out of the gas would increase substantially.

Galaxy interaction, if \src\ is an on-going merger, can be an
additional trigger of compressive gas motions.  As discussed in
\citet[][\S\,6.1]{Falkendal2018} and shown in Figure~\ref{fig:slit}, 
the dust continuum of \src\ has an elongated morphology which is different from the synthesized
beam, suggesting an ongoing merger. If the \CI\ spectrum of \src\
is interpreted as a two-component system rather than a rotating disk
(\S\ref{sec:CI}), it suggests that \src\ is undergoing a late-stage
merger that is prevalent among starbursting radio galaxies at high
redshift \citep{dBreuck2005,Ivison2012}.  In that case, \src\ may be
analogous to MRC\,0152-209 \citep{Emonts2015b}, wherein the timescale to
transfer all the gas is shorter than that of the gas outflow or orbital
timescales.  This timescale comparison would again be consistent with
a merger playing a key role in redistributing and likely compressing
the gas. Although we do not have a complete census of the sources of
momentum and energy in the gas of \src\ or any of the radio galaxies
in our various samples \citep[see, e.g.,][]{Dey1999, dBreuck2010,
Ivison2013,Emonts2014,Emonts2015b,Nesvadba2017,Falkendal2018},
it is clear from \src\ that both radio jets and the gravitational potential can 
induce compressive gas motions.
Deciding which one is more important depends on the timescales over which they have and can operate.

\subsection{Implications on the evolution of massive galaxies}\label{sec:discussion_gal}

\src\ is likely nearing the end of its star formation by depleting its molecular gas reservoir.
Although its current star-formation rate is high but perhaps less than it was during the past several 100 Myrs (see Sect.~\ref{subsubsec:UVcont}),
when compared to similarly massive galaxies,
in the long run, it will not significantly grow as the host galaxy is already massive.
This is true even when assuming 100\,\% efficiency in converting \Htwo\ to stars,
the available gas reservoir can only increase the stellar mass by $\sim$12\% at most.
In conjunction with star formation,
the active galactic nucleus helps to clear the galaxy of gas as evidenced by the fast, multi-phase outflow.
The obscured quasar slows down accretion onto \src\ during this ``blowout'' phase \citep{dMatteo2005,Hopkins2006},
eventually bringing the end to the star formation and blackhole growth in this massive galaxy when it runs of out fuel.

What, then, can we learn about the overall evolution of massive galaxies from this analysis of \src?
\citet{Falkendal2018} reported that most high-redshift radio galaxies (21 out of 25) are on their way to being quenched, meaning, their specific star formation rates are within, or below, those of "normal" galaxies at comparable masses and redshifts.
Only four radio galaxies of the sample, including the subject of this work, \src, have specific star-formation rates higher than ``normal'' galaxies.
At first glance,
these findings could be considered as evidence for feedback from AGN removing the gas and quenching star formation. 
The difficulty with this hypothesis, as \cite{Falkendal2018} found, is that the majority of sources had both powerful AGN and no apparent star formation (upper limits less than about 100-200 \Msun\ yr$^{-1}$).  
Of course, one could argue that in the case of short periods of time, AGN feedback is positive and then turns negative as it blows the gas away.  
However, that would mean that the dusty gas would have to be removed completely in less than a lifetime
of luminous radio sources ($\sim$10\,Myr), otherwise \citet{Falkendal2018} would have detected residual infrared emission. All the dust would have to be evacuated because any remaining dust would be heated by non-ionizing photons of the stellar population older than the radio source lifetime.

Given the difficulty with AGN feedback scenario, a more logical explanation is that the radio-loud AGN in these sources turn on during the ``final throes'' of the rapid growth of massive galaxies:
massive stars heat up and disperse gas,
creating favourable conditions in the ISM for AGN feedback to be efficient \citep{Biernacki2018}.
It is only when the gas fraction becomes relatively low that the feeding of the supermassive blackhole becomes significant \citep[see][]{Schawinski2009,Dubois2015,Volonteri2015b,Volonteri2015a}.
Starbursting radio galaxies therefore represents a brief but not unique phase in the life of massive galaxies,
as they transition toward radio-mode feedback wherein AGN prevents quiescent galaxies from gas cooling and star formation \citep{Croton2006,Best2014,Man2016a,Barisic2017}. The other, non-star forming radio galaxies have likely already entered this phase.

\section{Conclusions}\label{sec:conclusions}

Our spectroscopic analysis of \src\ has enabled us to obtain unprecedented constraints on the recent star-formation history and conditions of a massive, highly star-forming radio galaxy at $z=2.57$.
A plethora of absorption line features are detected in the deep \xs\ spectrum, 
including stellar photospheric and wind features indicative of OB-type stars,
in additional to emission lines and low-ionization absorption lines that are commonly seen in high-redshift star-forming galaxies.

The most significantly detected photospheric features, \SiII\,$\lambda$1485 and \SV\,$\lambda$1502, 
are used to constrain the nature of the recent star formation of \src\ through a comparison with \texttt{Starburst99} model spectra.
The star-formation history is inconsistent with a single burst or being continuous. 
Rather, more than one burst of star formation took place in the past $\sim100$\,Myr:
the most recent burst took place 4--7\,Myr ago explaining the presence of O- and early B-type stars,
and an older burst of at least 20\,Myr ago is needed to explain the presence of late B-type stars.
Evidence for the presence of stellar winds and weaker stellar absorption features in this spectrum corroborates these results.
A short star-formation timescale is further supported by the super-solar metallicity and alpha-element enhancement, 
as indicated by the presence and strengths of metal photospheric lines such as S, Si, O and Fe.

The photospheric lines are narrow and perhaps appear to have more than one velocity component.
Both of these properties are also seen in the ALMA \CI\ emission line tracing the cold, star-forming gas. 
This indicates that the stars formed in knots or clumps confined to small physical regions,
rather than dispersed throughout the host galaxy. 

The addition of ALMA \CI\ emission line of \src\ enables us to characterize the star-forming gas.
We derive a molecular gas mass of \MHtwo = $(3.9\pm1.0)\times10^{10}$\,\Msun,
which is only 12\% of its stellar mass.
The fraction of molecular-to-stellar mass is lower than galaxies at similar stellar mass and redshift.
Combining the gas mass with its high star-formation rate of $(1020^{+190}_{-170})$\,\Myr,
we estimate its star-formation efficiency to be $(26\pm8)$\,Gyr$^{-1}$,
or equivalently a gas depletion time of \tdepl\ = $(38\pm12)$\,Myr.
Its high star-formation efficiency is on par with the most efficiently star-forming galaxies at comparable mass and redshifts.
The surface density of the star-formation rate is as high as the dense clumps within molecular clouds and starburst galaxies,
significantly higher than that of disk galaxies.
These observations leads us to attribute the efficient star formation to gas compression,
which naturally explains the modest velocity dispersions ($\leqslant 55$\,\kms) of the young stars and of the cold gas.
Compressive gas flows can be triggered by radio jets and/or galaxy interaction.

The striking similarity between \src\ and other starbursting radio galaxies (e.g., 4C\,41.17, the Dragonfly and the Spiderweb) suggests that they may well represent a brief but not unique phase in the last active stage of massive galaxies.
Highly efficient star formation and AGN act together to deplete the cold gas reservoir as well as disperse it,
eventually quenching the star formation in these massive galaxies.

\begin{acknowledgements}

We wish to express our sincerest gratitude to the ESO data reduction
pipeline developers, especially Andrea Modigliani and Sabine Moehler from the user support department for their considerable help with the data
reduction. We thank Arjun Dey, Max Pettini, and Chuck Steidel for
making their spectra available for comparison,
and Guillaume Drouart for providing the
NIR photometry of this source, and all for discussions during the early
stages of this work. 
We thank the anonymous referee for helpful comments and criticisms that improved this study.
AM expresses her appreciation for the helpful discussions on stellar atmospheres with Rolf-Peter Kudritzki and Joachim Puls and for the useful conversations on a wide range of
topics with Federico Lelli, Johannes Zabl, Tina Peters, Sthabile Kowla, Zhi-Yu Zhang, Laura Zschaechner, James Matthews and Salvatore Cielo.  

Based on observations collected at the European Southern Observatory under ESO programme 092.B-0772(A).
This paper makes use of data from ALMA program ADS/JAO.ALMA\#2013.1.00521.S. 
ALMA is a partnership of ESO (representing its member states), NSF (USA) and NINS (Japan), together
with NRC (Canada) and NSC and ASIAA (Taiwan) and KASI (Republic of Korea),
in cooperation with the Republic of Chile. 
The Joint ALMA Observatory is operated by ESO, AUI/NRAO and NAOJ.
The Dunlap Institute is funded through an endowment established by the David Dunlap family and the University of Toronto.
\end{acknowledgements}

\bibliographystyle{aa} 
\bibliography{pks_SV} 

\begin{appendix}
\onecolumn

\section{Additional materials}

\input{z_em.tex}

\begin{figure}[!ht]
\includegraphics[width=0.9\linewidth]{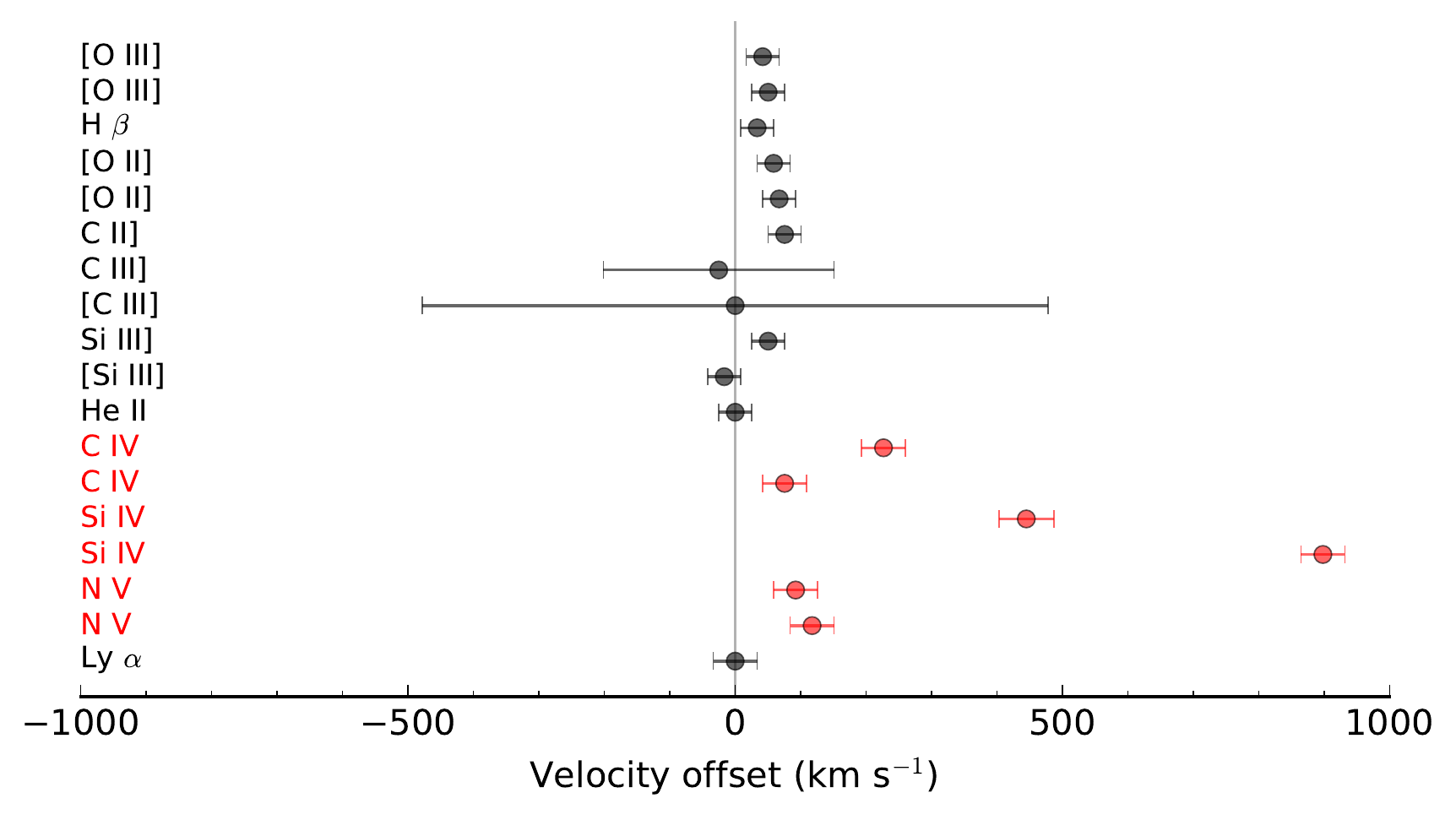}
\caption{Velocity offsets of emission lines in Table~\ref{table:redshifts} with respect to
\zsys\ = 2.5725 defined by the \HeII\ broad emission line.
The resonant doublets (in red) are affected by blueshifted absorption from the stellar winds in addition to the emission, explaining their redshifted velocities.}
\label{fig:em_velo}
\end{figure}

\newpage

\begin{figure*}[!ht]
\includegraphics[width=\linewidth]{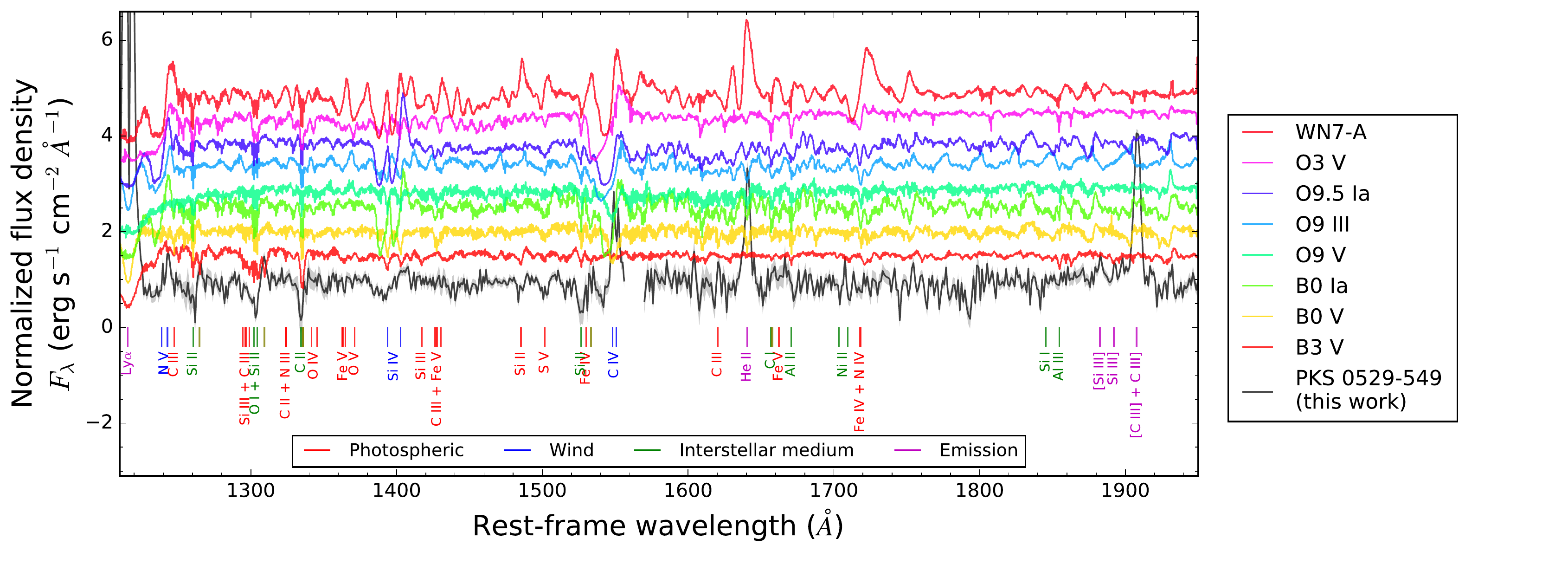}
\caption{
Rest-frame UV spectra of various types of O- and B-stars \citep{Walborn1985, Walborn1995b} drawn from the International Ultraviolet Explorer (IUE) Atlas\protect\footnotemark\ and that of \src.
At the top, we show the spectrum of a strong N-line Wolf-Rayet star, WN7-A, and as one moves down the plot, the stellar spectra are of stars of progressively later types from O3 main sequence stars to a B3 main sequence star (the specific types are indicated in the legend at the right).
For easier visualization, the stellar spectra have been smoothed by a Gaussian kernel and shifted upwards in increments of 0.5.
\label{fig:spec_OBstars}}
\end{figure*}

\footnotetext{\url{http://vizier.u-strasbg.fr}}

\begin{figure*}[!ht]
\includegraphics[width=\linewidth]{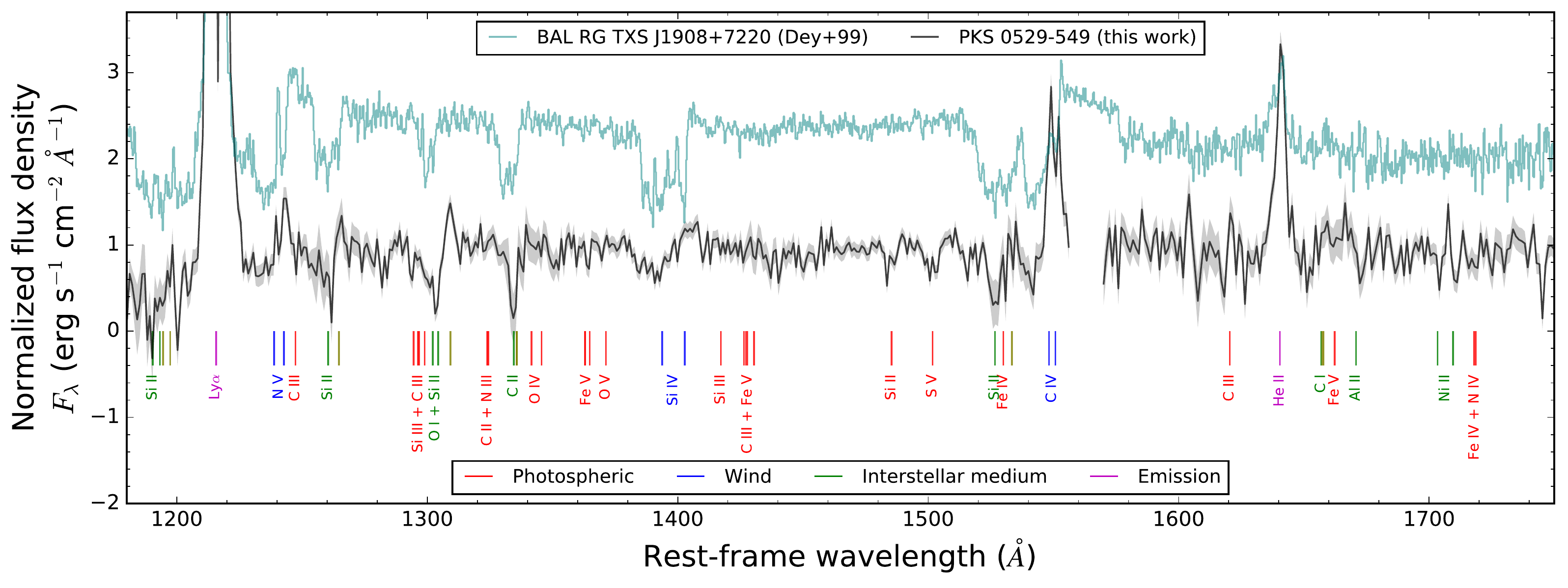}
\caption{Rest-frame UV spectrum of \src\ (this work) and TXS\,J1908+7220, a radio galaxy at $z=3.5356$ with broad absorption lines \citep{Dey1999, dBreuck2001}. The characteristics of the broad absorption lines in TXS\,J1908+7220 are generally consistent with those of BAL quasars. The spectrum of TXS\,J1908+7220 has been shifted upward by 1.3 for easier visualization.}
\label{fig:spec_bal}
\end{figure*}

\newpage

\begin{figure*}[!ht]
\begin{centering}
\includegraphics[width=0.75\linewidth]{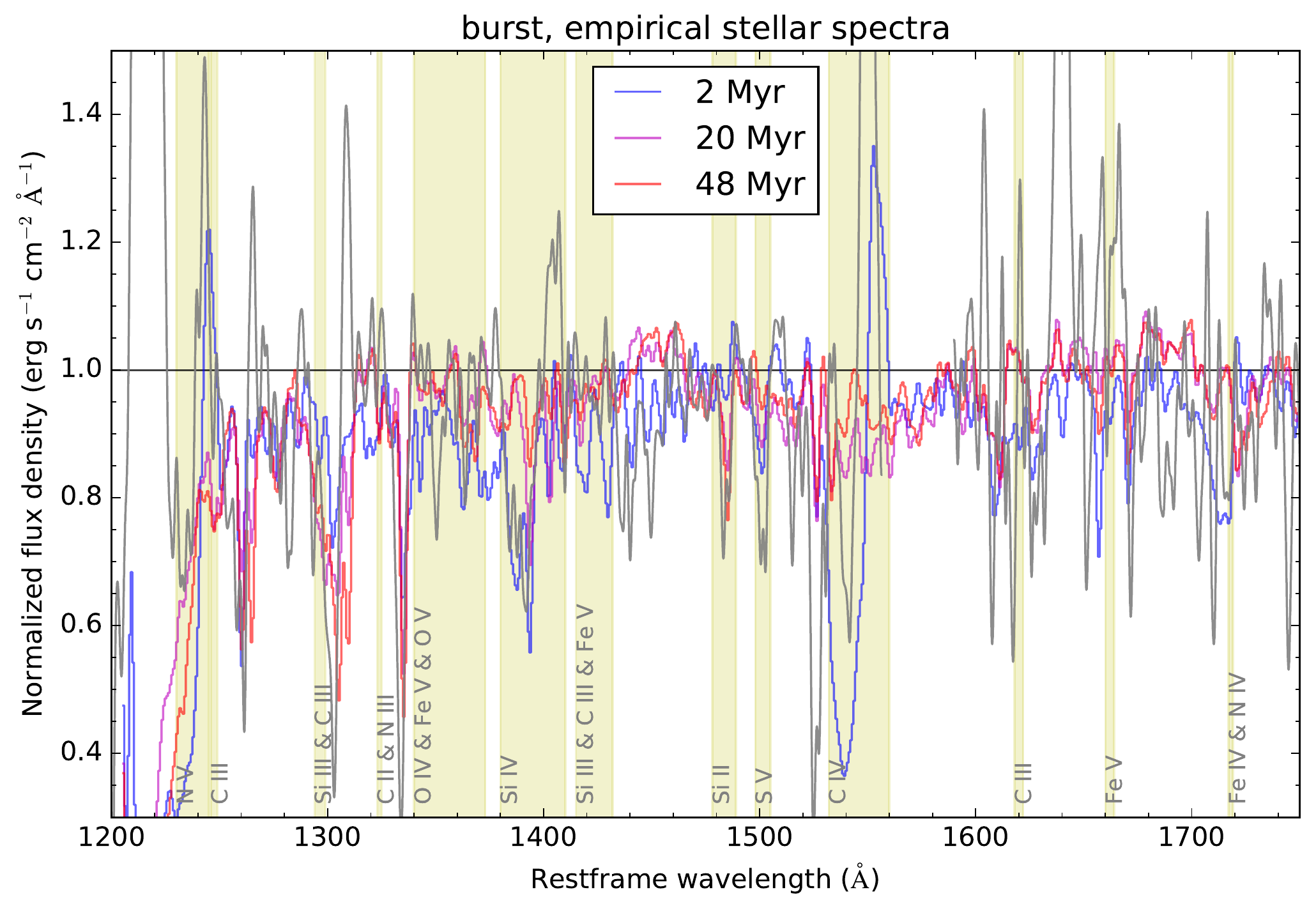}
\caption{The UV spectrum of a stellar population that experienced a single burst as it ages \citep{Leitherer1999}, 
compared with the observed spectrum of \src\ (grey line; this work) over the spectral region, 1200--1750\AA. We highlight
in yellow the spectral regions of containing possible photospheric signatures of young stars and stellar wind features. The times since the initial burst of individual spectral models are indicated in the legend on the top. The spectrum is normalized to the continuum and heavily smoothed for easier visualization.}
\label{fig:compare_sb99_instant}
\end{centering}
\end{figure*}

\begin{figure*}[!ht]
\begin{centering}
\includegraphics[width=0.75\linewidth]{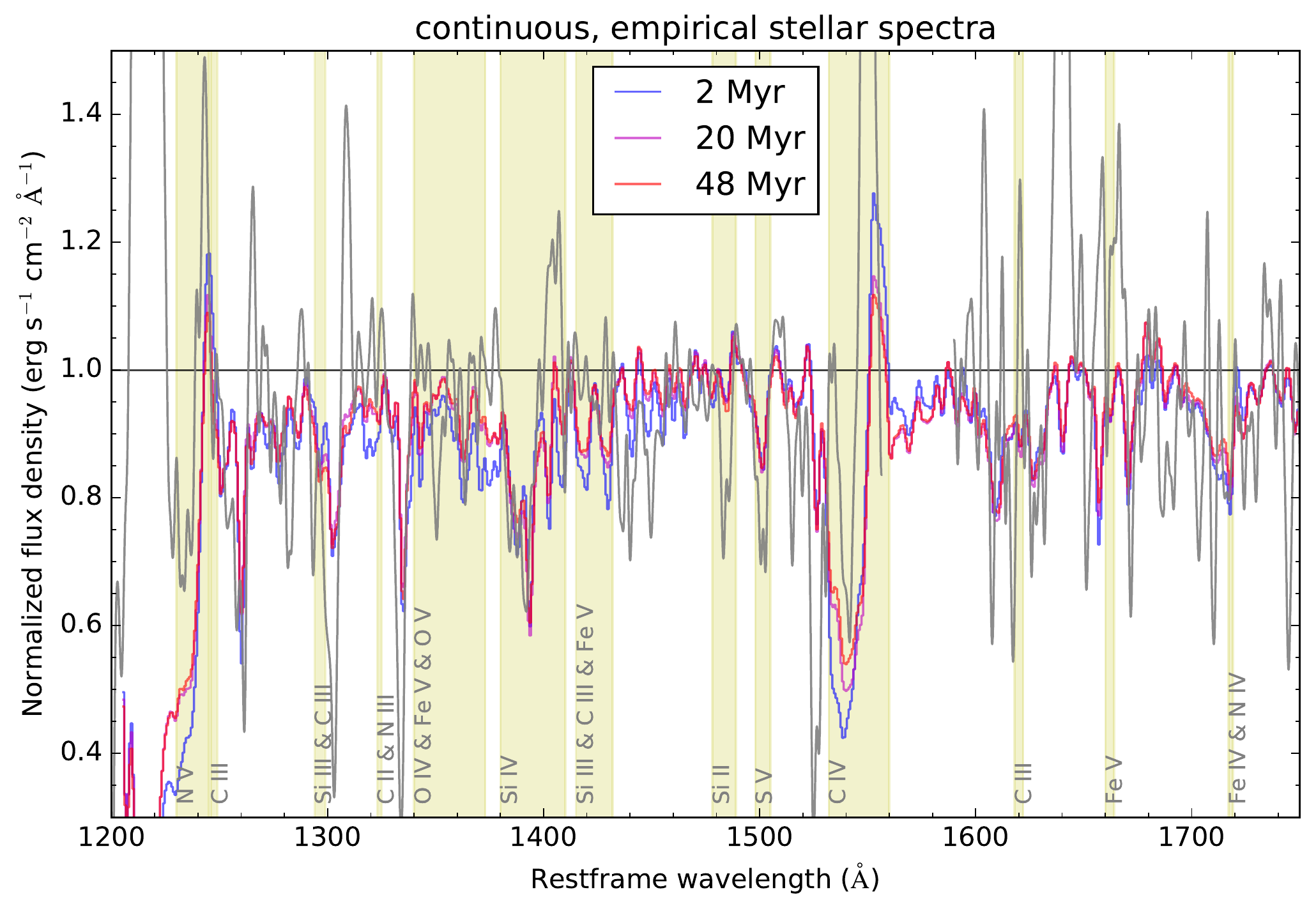}
\caption{Same as Figure~\ref{fig:compare_sb99_instant}, but for continuous star formation at a constant rate as a function of age.}
\label{fig:compare_sb99_const}
\end{centering}
\end{figure*}

\end{appendix}

\end{document}

%% file: table_photospheric.tex
\begin{table*}[!h]
\caption{Properties of absorption features with stellar photospheric origins.
The $\lambda_{\mathrm{rest,air}}$ column lists the rest-frame wavelengths of the features in air.
The $\lambda_{\mathrm{obs,air}}$ column lists the observed wavelengths as measured from the \xs\ spectrum of \src,
as inferred from fitting single or multiple Gaussian components to the absorption line profiles.
The $\sigma_{\mathrm{instrinsic}}$ column lists the intrinsic dispersion of the absorption lines, corrected for the instrument resolution.
Dagger symbols are used to indicate the lines possibly resolved into multiplets or multiple velocity components:
\SiII\,$\lambda$1485 (1484.87, 1485.22) and \SV\,$\lambda$1502 (1499.94, 1501.76, 1502.74, 1502.75, 1502.82).
As their intrinsic line ratios are undetermined,
we use the rest-frame wavelengths listed in \citet{dMello2000} to estimate their redshifts here.
\SiIII\,$\lambda$1294 is the bluest of a quadruplet of four photospheric
lines (\SiIII\,$\lambda\lambda\lambda$1294, 1296, 1298, and \CIII\,$\lambda$1296). However
the other three features lie in the broad absorption wing of the \OI\,$\lambda$1302.17 and \SiII\,$\lambda$1304.37 feature, and so we are unable to obtain accurate
redshift measurements.  \CII\,and \NIII\ is a blended absorption of \CII\,$\lambda$1323.93\,\& \NIII\,$\lambda$1324.32, both photospheric features.
We use their average rest-frame wavelength as indicated with $\star$.
}
\label{table:photospheric}
\begin{tabular*}{\linewidth}{c @{\extracolsep{\fill}} cccccc}
\hline
\hline
Feature & $\lambda_{\mathrm{rest,air}}$ & $\lambda_{\mathrm{obs,air}}$ & Redshift & $\sigma_{\mathrm{instrinsic}}$ & Comment \\
& (\AA) & (\AA) & & (km\,s$^{-1}$) & \\
\hline
\multicolumn{6}{c}{Single Gaussian component}\\
\hline
\CIII                & 1247.38 & 4455.68$\pm$0.3 & 2.5720$\pm$0.0004 & 17 & P-Cygni? \\
\SiIII               & 1294.54 & 4620.28$\pm$0.3 & 2.5690$\pm$0.0004 & 54 & Part of \SiIII\ and \CIII\ multiplet \\
\CII\,\&\,\NIII$\star$ & 1324.12 & 4724.63$\pm$0.3 & 2.5681$\pm$0.0006 & 29 & Blended \\
\OIV & 1341.64 & 4792.79$\pm$0.3 & 2.5723$\pm$0.0004 & 12 & \\
\SiIII & 1417.24 & 5060.85$\pm$0.3 & 2.5709$\pm$0.0004 & 11 & \\
\hline
\multicolumn{6}{c}{Multiple Gaussian components}\\
\hline
\SiII$\dagger$ & 1485.40 & 5298.74$\pm$0.1 & 2.5672$\pm$0.0004& 46 & Resolved into doublet \\
\SiII$\dagger$ & 1485.40 & 5308.09$\pm$0.1 & 2.5735$\pm$0.0004& 44 & \ditto\\
\SV$\dagger$   & 1501.76 & 5351.78$\pm$0.1 & 2.5637$\pm$0.0004 & 19 & Resolved into triplet \\
\SV$\dagger$   & 1501.76 & 5360.78$\pm$0.1 & 2.5697$\pm$0.0004 & 33 & \ditto \\
\SV$\dagger$   & 1501.76 & 5368.96$\pm$0.1 & 2.5751$\pm$0.0004 & 20 & \ditto \\
\HeII & 1640.42 & 5852.46$\pm$0.2 & 2.5677$\pm$0.0003 & 50 &  \\
\HeII & 1640.42 & 5857.07$\pm$0.1 & 2.5705$\pm$0.0002 & 25 & \\
\hline
\end{tabular*}
\end{table*}

%% file: table_vterm.tex
\begin{table}
\begin{centering}
\caption{Terminal velocities of resonant line absorptions estimated from their blue intercepts with the continuum, $\lambda_{\mathrm{edge}}$ quoted in the rest-frame.
Note that \NV\ absorption is filled by broad \La\ emission.
The errors are estimated as described in \S\ref{sec:winds}.}
\label{table:vterm}
\begin{tabular}{ccc}
\hline
\hline
Feature & $\lambda_{\mathrm{edge}}$ & \vterm \\
& (\AA) & (km\,s$^{-1}$) \\
\hline
\NV & $1226.3\pm1.8$ & $3000\pm400$ \\
\SiIV & $1380.1\pm1.7$ & $2900\pm400$ \\ 
\CIV & $1535.0\pm2.4$ & $2600\pm500$ \\
\hline
\end{tabular}
\end{centering}
\end{table}

%% file: table_CI.tex
\begin{table*}
\begin{centering}
\caption{Best-fitting parameters to the \CIlevels\ emission line as measured from the ALMA spectrum of \src. The systemic velocity is defined with respect to the \HeII\ emission (\S\ref{sec:sysred}). 
}
\label{table:CI}
\begin{tabular*}{\linewidth}{lccccccc}
\hline
\hline
Feature & Position & $\nu_{\mathrm{obs}}$ & Redshift & v-v$_{\mathrm{sys}}$ & FWHM & $S\Delta v$ & M$_{\mathrm{H_{2}}}$ \\
& & (GHz) & & (\kms) & (\kms) & (Jy\,km\,s$^{-1}$) & ($\times10^{10}$\Msun) \\
\hline
\CI\ red & $5^{h}30^{m}25^{s}.46 -54^{\circ}54'23\arcsec.14$ & 226.522$\pm$0.015 & 2.5729$\pm$0.0002 & $+34\pm20$ & 151$\pm$48 & 0.8$\pm$0.3 & $1.64\pm0.68$ \\
\CI\ blue & $5^{h}30^{m}25^{s}.41 -54^{\circ}54'23\arcsec.26$ & 226.780$\pm$0.017 & 2.5688$\pm$0.0003 & -307$\pm$23 & 202$\pm$54 & 1.1$\pm$0.4 & $2.23\pm0.79$ \\
Total & & & & & & 2.0$\pm$0.5 & $3.9\pm1.0$ \\
\hline
\end{tabular*}
\end{centering}
\end{table*}

%% file: table_hzrg.tex
\begin{table*}[htbp]
\begin{centering}
\caption{
Properties of efficiently star-forming, high-redshift radio galaxies. 
Star formation rates are derived from the \LIR\ (8 -- 1\,000\,\um) using the formula provided in \citet{Kennicutt1998a} assuming continuous bursts of age 10--100\,Myr, and converted to the \citet{Kroupa2001} IMF.
Molecular gas masses are estimated from either \CI\ or CO emission as described in \S\ref{sec:gas_depl}.
The quoted uncertainties refer to measurement errors.
The definitions of \fgas, SFE, and \tdepl\ can be found in \S\ref{sec:CI}.
$\Sigma_\mathrm{gas}$ is calculated as \MHtwo$/(\pi R^{2})$, where $R$ refers to the radius of the star-forming region.
$\Sigma_\mathrm{SFR}$ is calculated as SFR$/(\pi R^{2})$.
These numbers refer to the star formation within the host galaxies only, 
and do not include the \Htwo\ gas mass in the extended halo.
References: (a) \citealt{Falkendal2018}; (b) this work; (c) \citealt{Lelli2018}; (d) \citealt{Drouart2016}; (e) \citealt{dBreuck2005}; (f) \citealt{Miley1992}; (g) \citealt{Emonts2015a}; (h) \citealt{Emonts2015b}.
}
\label{table:hzrg_gas}
\begin{tabular}{lccccccccc}
\hline
\hline
 & SFR & \MHtwo & \fgas & SFE & \tdepl & size & log($\Sigma_\mathrm{gas}$) & log($\Sigma_\mathrm{SFR}$) \\
 & (\Myr) & ($10^{10}$\,\Msun)  & (\%)& (Gyr$^{-1}$) & (Myr) & (kpc) & log(\Msun\,pc$^{-2}$) & log(\Myr\,kpc$^{-2}$) \\
\hline
\src & $1020^{+190}_{-170}$ $^{a}$ & $3.9\pm1.0$ $^{b}$ & $12\pm7$ & $26\pm8$ & $38\pm12$ & 4 $^{c}$ & $2.9\pm0.1$ & $1.3\pm0.1$ & \\
4C\,41.17 & $2890^{+2880}_{-1980}$ $^{d}$ & $5.4\pm0.6$ $^{e}$ & $18\pm2$ & $54^{+54}_{-37}$ & $19^{+19}_{-13}$ & 4.3 $^{f}$ & $2.97\pm0.05$ & $1.7^{+0.3}_{-0.5}$ \\
MRC\,0152-209 & $1820^{+220}_{-270}$ $^{a}$ & $2.2\pm0.2$ $^{g}$ & $3.7\pm0.3$ & $83^{+12}_{-14}$ & $12\pm2$ & 1.5 $^{h}$ & $3.49\pm0.04$ & $2.41^{+0.05}_{-0.07}$ \\
\hline
\end{tabular}
\end{centering}
\end{table*}

%% file: z_em.tex
\begin{table}[h]
\begin{centering}
\caption{Redshift measurements of \src\ based on strongest emission lines detected in the \xs\ spectrum.
$\dagger$ indicates the feature used to define the systemic redshift.
Listed wavelengths are as measured in air,
along with the measurement errors resulting from the line-fitting.
The redshift uncertainties include the listed measurement errors of the observed wavelengths,
as well as systematic errors due to the wavelength calibration, spectral resolution and uncertainties in the theoretical rest-frame wavelengths.
We note possible anomalies in the redshift measurements from these lines, including known and possible line blending, presence of absorption, and obscuration by bright sky emission lines.
The large redshift uncertainties of the \CIIIsf\ and \CIIIf\ doublet redshifts are due to blending: determining the doublet line ratio requires knowledge of the electron density \citep{Ferland1981,Keenan1992}.
The \OIIIsf\,$\lambda\lambda$1661, 1666 are on top of skylines and redshift measurements are not possible.
}
\label{table:redshifts}
\begin{tabular*}{\linewidth}{c @{\extracolsep{\fill}} cccc}
\hline
\hline
Feature & $\lambda_{\mathrm{rest,air}}$ & $\lambda_{\mathrm{obs,air}}$ & Redshift & Comments \\
& (\AA) & (\AA) & & \\
\hline
\La      & 1215.67 & 4343.02$\pm$0.05 & 2.5725$\pm$0.0004 & Fitted simultaneously with a narrow absorption component \\
\NV      & 1238.82 & 4427.42$\pm$0.17 & 2.5739$\pm$0.0004 & Affected by blueshifted absorption\\
\NV      & 1242.80 & 4441.34$\pm$0.21 & 2.5736$\pm$0.0004 & \ditto \\
\SiIV    & 1393.76 & 4994.13$\pm$0.43 & 2.5832$\pm$0.0004 & \ditto \\
\SiIV    & 1402.77 & 5018.79$\pm$0.70 & 2.5778$\pm$0.0005 & \ditto \\
\CIV     & 1548.20 & 5532.30$\pm$0.10 & 2.5734$\pm$0.0004 & \ditto \\
\CIV     & 1550.78 & 5544.26$\pm$0.10 & 2.5752$\pm$0.0004 & \ditto \\
\HeII$\dagger$    & 1640.42 & 5860.46$\pm$0.14 & 2.5725$\pm$0.0003 & Asymmetric; Fig.~\ref{fig:vel_comp} \\
\SiIIIf & 1882.47 & 6724.81$\pm$0.26 & 2.5723$\pm$0.0003 & Possible blend of 4 \SiIIIsf\ lines \\
\SiIIIsf & 1892.03 &  6760.34$\pm$0.45 & 2.5731$\pm$0.0003 &  \\
\CIIIf  & 1906.68 & 6811.70$\pm$15.19 & 2.5725$\pm$0.0057 & \CIIIsf\ doublet blended; Skyline\\
\CIIIsf  & 1908.73 & 6818.442$\pm$5.66 & 2.5722$\pm$0.0021 & \ditto \\
\CIIsf & 2325.40 & 8309.612$\pm$0.36 & 2.5734$\pm$0.0003 & Possible blend of 4 \CIIsf\ and \OIII\ lines; Skylines \\
\OII     & 3726.03 & 13314.33$\pm$0.64 & 2.5733$\pm$0.0003 & \OII\ doublet blended; Skyline \\
\OII     & 3728.82 & 13323.71$\pm$0.96 & 2.5732$\pm$0.0003 & \ditto \\ 
\Hb      & 4861.33 & 17369.18$\pm$0.96 & 2.5729$\pm$0.0003 & Skylines \\
\OIII    & 4958.91 & 17718.74$\pm$0.96 & 2.5731$\pm$0.0003 & Skylines\\
\OIII    & 5006.84 & 17889.57$\pm$0.47 & 2.5730$\pm$0.0003 & Skylines \\
\hline
\end{tabular*}
\end{centering}
\end{table}